\documentclass[a4paper,12pt]{article}
\usepackage{graphicx}
\usepackage{multirow}
\usepackage{bbm}
\usepackage[labelfont=bf,font=normalsize]{caption}
\usepackage{float}
\usepackage{footnote}
\usepackage{amssymb}
\usepackage{pifont}
\usepackage{ifpdf}
\ifpdf
\else
\usepackage{pstcol,pst-fill,pstricks}
\fi
\usepackage{natbib}
\usepackage{mathpazo}
\usepackage{import}
\usepackage{dsfont}
\usepackage{enumerate}
\usepackage{lscape}
\usepackage{epsfig}
\usepackage[latin1]{inputenc}
\usepackage[OT1]{fontenc}
\usepackage[T1]{fontenc}

\usepackage[margin=1in]{geometry}
\usepackage[doublespacing]{setspace}

\usepackage{color}
\usepackage{mdframed}
\usepackage{tikz}
\usepackage{tkz-graph}
\usepackage[bottom,flushmargin]{footmisc}
\usepackage{subcaption}  
\usepackage{times}
\usepackage{fancyhdr,graphicx,amsmath,amssymb}
\usepackage[ruled,vlined]{algorithm2e}
\usepackage{mathtools}
\usepackage{amsmath}
\usepackage{amsthm}
\usepackage{thmtools}
\usepackage{calrsfs}
\usepackage{booktabs}

\makeatletter
\newcommand*\bigcdot{\mathpalette\bigcdot@{.5}}
\newcommand*\bigcdot@[2]{\mathbin{\vcenter{\hbox{\scalebox{#2}{$\m@th#1\bullet$}}}}}
\makeatother
\usepackage{authblk}
\usepackage{setspace}
 \newcommand{\be}{\begin{equation}}
\newcommand{\ee}{\end{equation}}
\newcommand{\E}{\mathbb{E}}
\newtheorem{assumption}{Assumption}
\newcommand{\pbar}{\overline{p}}
\newcommand{\dbar}{\overline{d}}

\newcommand{\pibar}{\overline{\pi}}
\newcommand{\cbar}{\overline{c}}

\DeclareMathAlphabet{\pazocal}{OMS}{zplm}{m}{n}
\allowdisplaybreaks

\usepackage[hyperfootnotes=false]{hyperref}
 \hypersetup{colorlinks=true,
 linkcolor=blue,
 citecolor=blue,
 filecolor=black,
 urlcolor=black,
 pdfstartview={Fit},
pdfpagemode=UseNone
 }
\usepackage{cleveref}

\declaretheoremstyle[
  headfont=\normalfont\scshape,
  numbered=unless unique,
  bodyfont=\normalfont,
  spaceabove=1em,
  spacebelow=1em,
]{exmpstyle}

\newcommand{\defeq}{\vcentcolon=}

\def\ee{\mathsf{e}}

\DeclareMathOperator{\tr}{tr}

\usepackage[mathlines]{lineno}

\newtheorem{prop}{Proposition}

\theoremstyle{definition}
\newtheorem{definition}{Definition}[section]

\newtheorem{corollary}{Corollary}

\title{A Step Towards a Solution to the Confidence-Driven Liquidity Trap Morass\thanks{Corresponding author: Jordan Roulleau-Pasdeloup: \href{mailto:jordan.roulleau@gmail.com}{jordan.roulleau@gmail.com}. We thank Jean-Baptiste Michau, Olivier Loisel, Ivan Shchapov, Boragan Aruoba and an anonymous referee as well as seminar participants at the National University of Singapore and the \'Ecole Polytechnique for insightful comments.}}
\author[1]{Haochun Ma}
\author[2]{Jordan Roulleau-Pasdeloup}
\affil[1]{National University of Singapore}
\affil[2]{Banque de France}
\date{\today}

\begin{document}
\begin{titlepage}
\maketitle
\thispagestyle{empty}

\abstract{Depending on the persistence of a one-off shock bringing the economy to the Effective Lower Bound (ELB), the standard New Keynesian model predicts starkly different conclusions. We offer a potential solution to this morass by assuming that the one-off shock is such that the economy necessarily leaves the ELB in finite time. Under these assumptions, we prove that the confidence-driven liquidity trap \`a la \cite{mertens2014fiscal} doesn't arise as an equilibrium outcome and that there is a unique path back for a given steady state. Furthermore, under our shock specification the effect of government spending on consumption is necessarily positive and does not switch signs but may grow unbounded \textemdash a puzzle.}
\vspace{.5cm}\\
\noindent{\bfseries JEL Codes: E32, E43, E52, E63} \\
\noindent{\bfseries Keywords: Truncated Markov Chain, Effective Lower Bound, Fiscal Policy} \\
\end{titlepage}


\section{Introduction}

With the Great Recession and the Covid-19 crises, advanced economies have experienced two long-lasting episodes where the Central Bank's interest rate was stuck at its Effective Lower Bound \textemdash henceforth ELB. Given the secular decline in interest rates that have been observed across the board for the same countries, they are bound to experience it again sometime in the near future. Against this backdrop, do we have a reliable framework to think about the effects of policy at the ELB?

In this paper, we argue that the standard New Keynesian (NK) model with the standard shock structure that is often used to understand the mechanisms of policy transmission at the ELB may not be a reliable framework. To understand why, note that ELB episodes in the standard NK model are often modeled as a result of shocks that follow a standard 2-state absorbing Markov chain where the absorbing state is the intended steady state.\footnote{Following the seminal contribution of \cite{Benhabib2001perils}, we know that there exists a second, unintended, steady state which features a binding Effective Lower Bound and below target inflation. Throughout most of the paper we will assume that the economy returns to the intended steady state but we will also explore what happens if the economy ends up returning to the unintended steady state.} Roughly speaking, this shock structure is the Markov Chain equivalent of a standard "MIT shock" (\cite{Boppart2018exploiting})\footnote{More general Markov chains in which the state where the ELB doesn't bind isn't absorbing but recurring instead include \cite{Aruoba2018} and \cite{Nakata2022expectations}.} . This shock structure allows one to replicate occasional, but long-lasting ELB episodes as argued in \cite{Dordal}. Let us denote by $1-p\in [0,1]$ the probability that this shock returns to its absorbing state every period. \cite{eggertsson2011fiscal} shows that there exists a threshold probability $\overline{p}$ such that if $0<p<\overline{p}$ then one can end up in a fundamentals-driven ELB episode with a sufficiently large negative demand shock. Assuming a perfectly correlated government spending shock, he proves that such a policy \textit{crowds private consumption in} at the ELB. Using a similar setup but assuming that $\overline{p}< p<1$ instead, \cite{mertens2014fiscal} show that one has two Minimum State Variable (MSV) equilibria, one of which features a confidence-driven ELB episode. They interpret this as a sunspot equilibrium and show that a perfectly correlated government spending shock \textit{crowds private consumption out}. In addition, \cite{Ascari2022unbearable} show that in that situation, if the underlying shock is too large, no MSV equilibrium exists.

Accordingly, the main message from the existing literature is that the effects of government spending can flip signs depending on the source of the ELB episode. In a related paper in which they study the fiscal multiplier in a rich set of medium scale models without an ELB constraint, \cite{Leeper2017clearing} refer to this state of affairs as a \textit{morass}. Therefore, we argue that the existing literature on fundamentals- vs confidence-driven liquidity traps features such a morass. Given that the confidence-driven part produces perhaps the most counter-intuitive\footnote{By counter-intuitive we mean that, for example, government spending crowds out consumption with a fixed nominal interest rate. Indeed, one would expect higher inflation and lower interest rates after an increase in government spending, which would crowd consumption in.} results, we name it the \textit{Confidence-Driven
Liquidity Trap Morass}.

In this paper, we show that truncating the underlying Markov chain is a step towards solving this morass. Namely, it gives rise to a framework with consistent policy recommendations with no sign flips. Following \cite{eggertsson2003optimal}, we assume that there exists a maximum time period $\ell$ after which the Markov chain has to return to its absorbing state. This shock structure will have a number of interesting properties. First, we show that under such a truncated Markov chain there exists only one equilibrium.\footnote{This uniqueness result holds whether the economy is assumed to revert to the intended or unintended steady state after the shock subsides.} As a result, non-existence of an equilibrium never arises in our framework.  Therefore, one can consider policy experiments that are well-defined regardless of the degree of persistence or the size of the shock. That unique equilibrium may be unstable, by which we mean that the impact effect on, say, inflation may grow unbounded as the truncation horizon $\ell$ goes to $\infty$. Whenever it is stable, our equilibrium construction converges to the one in \cite{eggertsson2011fiscal} as $\ell\to\infty$. The parameter region for which our equilibrium is unstable corresponds to the one that leads to a confidence-driven liquidity trap in \cite{mertens2014fiscal}, which then doesn't arise in our framework. Note that this result rests on the assumptions of a one-off, truncated Markov Chain shock in a standard New Keynesian model and doesn't imply that confidence-driven liquidity traps aren't pervasive in (possibly more general) New Keynesian models with a different shock structure. See for example \cite{Benhabib2001perils}, \cite{Schmitt2017liquidity}, \cite{Armenter2018perils}, \cite{Aruoba2018}, \cite{Benigno2018stagnation}, \cite{Heathcote2018wealth}, \cite{Nakata2022expectations}, \cite{Caramp2023bond}  
as well as \cite{Cuba2024understanding} for such examples.

Zooming in on fiscal policy, we find that, regardless of the persistence level and whether the economy converges back to the intended or unintended steady state, consumption is \textit{crowded in}, provided the ELB is long enough. This is in stark contrast with the conclusions reached in \cite{mertens2014fiscal} for a similar model. In our framework, one can compute in closed form the impact multiplier effect of government spending on, say, consumption as a function of $\ell$. It turns out that this effect obeys a simple second order recursion of the form $m(\ell) = a\cdot m(\ell-1)+b\cdot m(\ell-2) + c^*_m $, where $m(\ell)$ denotes the multiplier effect for a maximum ELB duration of $\ell$, the coefficients $a,b$ are functions of structural parameters and $c^*_m$ is given by $(1-a-b)$ times the consumption multiplier that obtains with the methods developed in \cite{eggertsson2011fiscal}, \cite{mertens2014fiscal}. Note that the initial conditions of this recursion are given by the pair $m(1), m(2)$: the effects under a shock that can last at most 1 and 2 periods respectively. We show that the solution studied in \cite{mertens2014fiscal} is indeed a fixed point of this recursion. However, the assumption of $\overline{p}< p<1$ coupled with a large enough shock precludes the recursion from converging to its fixed point: it diverges away instead. Therefore, under the assumption that $\overline{p}< p<1$, the fixed point is unstable.\footnote{We thank an anonymous referee for suggesting us to frame it in this way.} If one assumes $0<p<\overline{p}$ instead, as $\ell\to\infty$ the recursion converges to its fixed point, which in turn coincides with the constant equilibrium constructed in \cite{eggertsson2011fiscal}. 

Mathematically, one can solve for the recursion and express $m(\ell)$ for a given $\ell$ as a partial geometric sum with persistence $p$ as its main argument. For a large enough demand shock, the coefficients $a$ and $b$ will reflect dynamics at the ELB. In that situation, it turns out that the radius of convergence for this geometric series as $\ell\to\infty$ is given by $0<p<\pbar$. In this context, the equilibrium/multiplier constructed in \cite{mertens2014fiscal} reflects the \textit{analytical continuation} of the geometric series. Strictly speaking, the geometric series is not defined for $p>\pbar$, but its analytical continuation is.\footnote{As a simple example, the geometric series $f(x) = \sum_{k=0}^{\infty}x^k$ has a radius of convergence of $|x|<1$ meaning that $\sum_{k=0}^{\infty}x^k = 1/(1-x)$ iff $|x|<1$. For $|x|>1$, the geometric series isn't well defined anymore but $1/(1-x)$ still is, except for $x=1$. In that sense, $1/(1-x)$ is the analytical continuation of $f(x)$.} 

Finally, our approach also sheds some light on whether the NK model delivers the well known puzzles (see \cite{Cochrane2017new},  \cite{Michaillat2021resolving}) at the ELB. By a puzzle we mean that the impact effect on, say, inflation, grows unbounded as the duration of the ELB episode increases. Indeed, it has been shown (see \cite{bilbiie2022neo}) that the NK model displays a bifurcation at the ELB around $\overline{p}$ where it produces unbounded outcomes. Using our method, we show two things. First, $p>\overline{p}$ is a \textit{necessary} but not sufficient condition for a puzzle to arise. We need the underlying demand shock to be large enough in magnitude in addition. If the demand shock is not large enough, then there is no puzzle regardless of the level of persistence. Second, a puzzle arises for a much larger region of the parameter space. Let us denote by $\overline{d}(p)$ the minimum value of the demand shock such that the model ends up at the ELB eventually. While in the cited literature a puzzle arises in the vicinity of $\overline{p}$, in our case it arises in a subset of $[\overline{p}\ \ 1]\times [\overline{d}(p)\ \ d^{max}]$, where $d^{max}$ is the maximum value of the shock. We quantify the size of these two regions using a calibrated example later on in the main text.

\textbf{Related Literature}\textemdash Given our focus on the effects of policy at the ELB, this paper is related to \cite{eggertsson2011fiscal}, \cite{christiano2011government}, \cite{woodford2011simple}, \cite{mertens2014fiscal}, \cite{Schmidt2017fiscal}, \cite{Cochrane2017new},  \cite{wieland2018state}, \cite{Hills2018fiscal}, \cite{Miyamoto2018}, \cite{wieland2019zero}, \cite{Nakata2022expectations}, \cite{bilbiie2022neo},  \cite{Woodford2022fiscal} and \cite{Holden2023existence}. In particular, we are interested in the existence of a confidence-driven ELB episode that has been studied in \cite{Benhabib2001perils}, \cite{mertens2014fiscal}, \cite{Schmitt2017liquidity}, \cite{Armenter2018perils}, \cite{Aruoba2018}, \cite{Benigno2018stagnation}, \cite{Heathcote2018wealth}, \cite{bilbiie2022neo}, \cite{Nakata2022expectations}, \cite{Ascari2022unbearable}, \cite{ascari2023coherence}, \cite{Caramp2023bond}, \cite{Cuba2024understanding} and \cite{Murakami2026restoring}. 

To construct the solution of our model, we borrow from the method developed in \cite{eggertsson2003optimal} and more recently \cite{eggertsson2021toolkit}. More specifically and as alluded to before, we rely on an absorbing Markov chain that is truncated in the sense that it is forced to go back to its absorbing state in finite time. As a result, the framework that we use shares some similarities with the one introduced in \cite{woodford2019monetary} in which firms and consumers can only plan for a finite horizon. Typically, this is coupled with an assumption about learning in a backward fashion beyond the planning horizon. This setup has been further studied in \cite{gust2022short,gust2024inflation} as well as \cite{dupraz2025keeping}. The assumption of backward learning injects endogenous persistence and thus precludes the derivation of closed form results for this class of models. In contrast, the assumption of truncated Markov chains that we rely on allows us to fully characterize the dynamics of the model in closed form. As a result, our framework should be viewed as a complement to \cite{Woodford2022fiscal}, who use \cite{woodford2019monetary}'s finite horizon planning framework without learning: in their case, the effects of the ELB binding percolate through short run expectations only for a finite number of periods because of the finite planning horizon whereas for us it is because of the finite shock duration. In addition, note that the case where $\pbar<p<1$ is central to our analysis while it is explicitly ruled out in \cite{Woodford2022fiscal} \textemdash see their equation (15). Our method is also related to the one developed in \cite{Holden2023existence}, in which a deterministic economy is assumed to go escape the ELB in finite time. The main difference is that \cite{Holden2023existence} focuses on uniqueness while we also focus on the stability of the solution. Our assumption is also less demanding regarding the agent's expectations in that they don't need to know the exact date at which the economy gets out of the ELB, only that it has to exit by a certain date. This leaves open the possibility that the economy randomly exits before that time period.\footnote{In addition, given the auto-regressive nature of the shocks considered in \cite{Holden2023existence}, one typically needs a bigger shock to generate a longer ELB episode. This may cause a counterfactually large decline in inflation on impact. In the case of a Markov chain, one can generate long spells at the ELB with a moderate initial decline in inflation provided that the chain stays in the low state for a long period in a given draw.}

Our shock specification leads us to conclude that, for a standard New Keynesian model with a one-off Markov shock structure, the expectations-driven episode only happens when $\ell=+\infty$ and that the range of parameters that yield a puzzling behavior is much larger than initially thought. Given our focus on these puzzles, we are also related to \cite{eggertsson2010paradox}, \cite{carlstrom2015inflation}, \cite{Cochrane2017new}, \cite{Michaillat2021resolving}, \cite{diba2021pegging}, \cite{Del2023forward}, \cite{gibbs2023does} and \cite{eskelinen2024resolving}. 

This paper is structured as follows. In Section \ref{ref:picturing}, we illustrate the morass graphically for a standard New Keynesian model. In Section \ref{sec:resolving_morass}, we prove formally and illustrate graphically how the assumption of a truncated Markov chain clears up the morass in that specific context. In Section \ref{sec:policies}, we use our framework to study the effects of fiscal policy at the ELB. We finally conclude in Section \ref{sec:conclusion}.

\section{Picturing the Morass}
\label{ref:picturing}

In order to illustrate the morass both intuitively and graphically, we borrow the framework from \cite{ascari2023coherence}. To facilitate the discussion of the effects of government spending later on, we write it in terms of private consumption. This setup nests the conventional three-equation New Keynesian model as a special case and allows for a class of behavioral models by relaxing the assumption of full-information rational expectations. The model is described by the following equations:
\begin{align}
\label{eq:Euler}
    c_t &= m_{xx} \mathbb{E}_t c_{t+1} - \sigma(i_t - m_{x\pi} \mathbb{E}_t \pi_{t+1} + d_t) \\
    \label{eq:NKPC}
    \pi_t &= m_{\pi\pi} \beta \mathbb{E}_t \pi_{t+1} + \kappa(1+\eta \cbar) c_t + \kappa \eta (1-\cbar) g_t \\
    \label{eq:Taylor}
    i_t &= \max \{\psi \pi_t, -\mu \}
\end{align}
where \(c_t\) denotes private consumption, \(i_t\) the nominal interest rate, and \(\pi_t\) the inflation rate, while $d_t$ is a standard liquidity preference shock and $g_t$ is a government spending shock. We assume that $m_{ij}\in (0,1]$ for $i,j\in\left\{x,\pi\right\}$. In this context, setting \(m_{xx} = m_{x\pi} = m_{\pi\pi} = 1\) recovers the standard New Keynesian model. Further,  $\mu$ is the real interest rate in the steady state with zero inflation. It then follows from equation \eqref{eq:Taylor} that the ELB will be a binding constraint whenever inflation is such that $\pi_t\leq \overline{\pi}\defeq -\mu/\psi$, where $\psi$ governs how aggressively the Central Bank reacts to inflation. Following \cite{eggertsson2011fiscal}, we assume for now that the shocks $d_t$ and $g_t$ both follow absorbing two-state Markov chains governed by the following initial distribution $U$,  transition matrix $\mathbf{P}$ and vector of states $S$:
\begin{align*}
U = 
\begin{bmatrix}
1 & 0    
\end{bmatrix}
\quad\mathbf{P} = \begin{bmatrix}
p & 1 - p \\
0 & 1
\end{bmatrix}
\quad
S = \begin{bmatrix}
s & 0    
\end{bmatrix}
^\top,
\end{align*}
where $s=d$ for the demand shock and $s=g$ for the government spending shock. For now and until section \ref{sec:policies}, we focus on the effects of the demand shock and we assume for simplicity that $g=0$. We use an uppercase bold font to represent matrices and uppercase to represent vectors while scalars are represented with a lowercase letter.
Following most of the literature, we maintain the following assumption throughout the paper:
\begin{assumption}
\label{ass:intended_SS}
The economy returns to its intended steady state with $\pi=0$ and $c=0$ in the long run.    
\end{assumption}
Indeed, it has been well-known at least since \cite{Benhabib2001perils} that there exists a second steady state with deflation and zero interest rates. In that sense, we follow \cite{eggertsson2011fiscal} \textemdash see the discussion after Proposition 2, page 71. \cite{Woodford2022fiscal} make a similar assumption by specifying their terminal value function with inflation being on target \textemdash see their equation (5). A related assumption can also be found in \cite{Holden2023existence} for the same model under perfect foresight. In contrast, \cite{Ascari2022unbearable} explicitly consider the possible return to a long-run steady state with a binding ELB. In this context, our contribution will be to show that there exists a unique path back to the intended steady state under our shock formulation, while in \cite{Ascari2022unbearable} there can be 0 or even 2. These possibilities never arise in either \cite{eggertsson2011fiscal} or \cite{Woodford2022fiscal}. Note that Assumption \ref{ass:intended_SS} is made without loss of generality and we will show later that the main result of this section (the equilibrium path back towards steady state being unique) also applies if the economy is assumed to return to the unintended steady state instead.

As has become standard in this literature, we first guess a MSV equilibrium solution where $c_t$ and $\pi_t$ can be written as a linear function of the shock $d_t$ and then verify if/when the ELB is binding or not. In practice, this amounts to work with a model where we can write $\E_t\pi_{t+1} = p\cdot \pi_t$ and likewise for $\E_t c_{t+1}$. Using this, one can compute the aggregate supply/demand curves of this economy conditional on the initial realization $d$. The main question that we seek to answer at this point is: for given values of $p,d$, how many MSV paths\footnote{With a slight abuse of language, given our assumption \ref{ass:intended_SS} and in the context of the standard New Keynesian model, when we say that there is an MSV path back to the intended steady state we mean that this path is constructed with an invariant MSV equilibrium that only consists in the exogenous shocks and involves no extra persistence beyond the one coming from the shock themselves\textemdash see \cite{bilbiie2022neo}. In practice, under the absorbing Markov chain assumption that path will be characterized by a step function.} back to the intended steady state are there? We answer that question in the following proposition.

\begin{prop}
\label{prop:existence_msv_equilibrium}
For a vector $[p\ \ d]$ in $[0,1]\times[0,d^{max}]$, there can be 0, 1 or 2 MSV equilibria. Further, there exists a pair of thresholds $\overline{p}$ and $\overline{d}(p)$ such that:
\begin{enumerate}
    \item If $p<\overline{p}$, then there exists only one MSV equilibrium (the ELB binds if $d \geq \bar{d}(p)$, otherwise it does not).
    \item If $p>\overline{p}$ and $d<\overline{d}(p)$, then there are two MSV equilibria: one where the ELB binds, and one where it does not.
    \item If $p>\overline{p}$ and $d>\overline{d}(p)$, there does not exist any MSV equilibrium.
\end{enumerate}
\end{prop}
\begin{proof}
This proposition essentially summarizes results that have been established in the existing literature and as a result the proof is relegated to the online Appendix, section A.   
\end{proof}

\begin{figure}[!htb]
  \centering
  \begin{subfigure}[t]{0.3\textwidth}
    \centering
    \includegraphics[width=\textwidth]{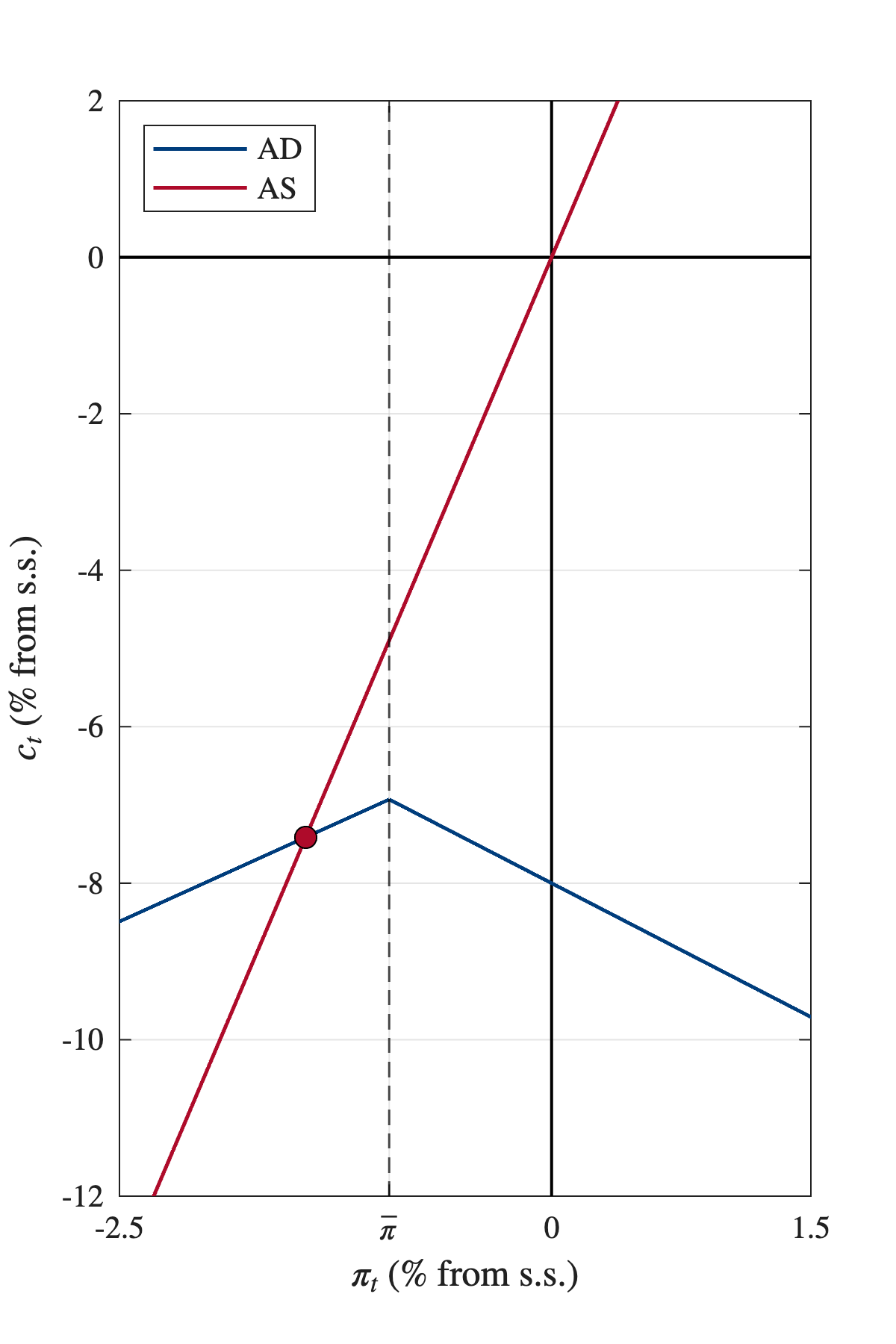}
    \caption{Case 1}
  \end{subfigure}\hfill
  \begin{subfigure}[t]{0.3\textwidth}
    \centering
    \includegraphics[width=\textwidth]{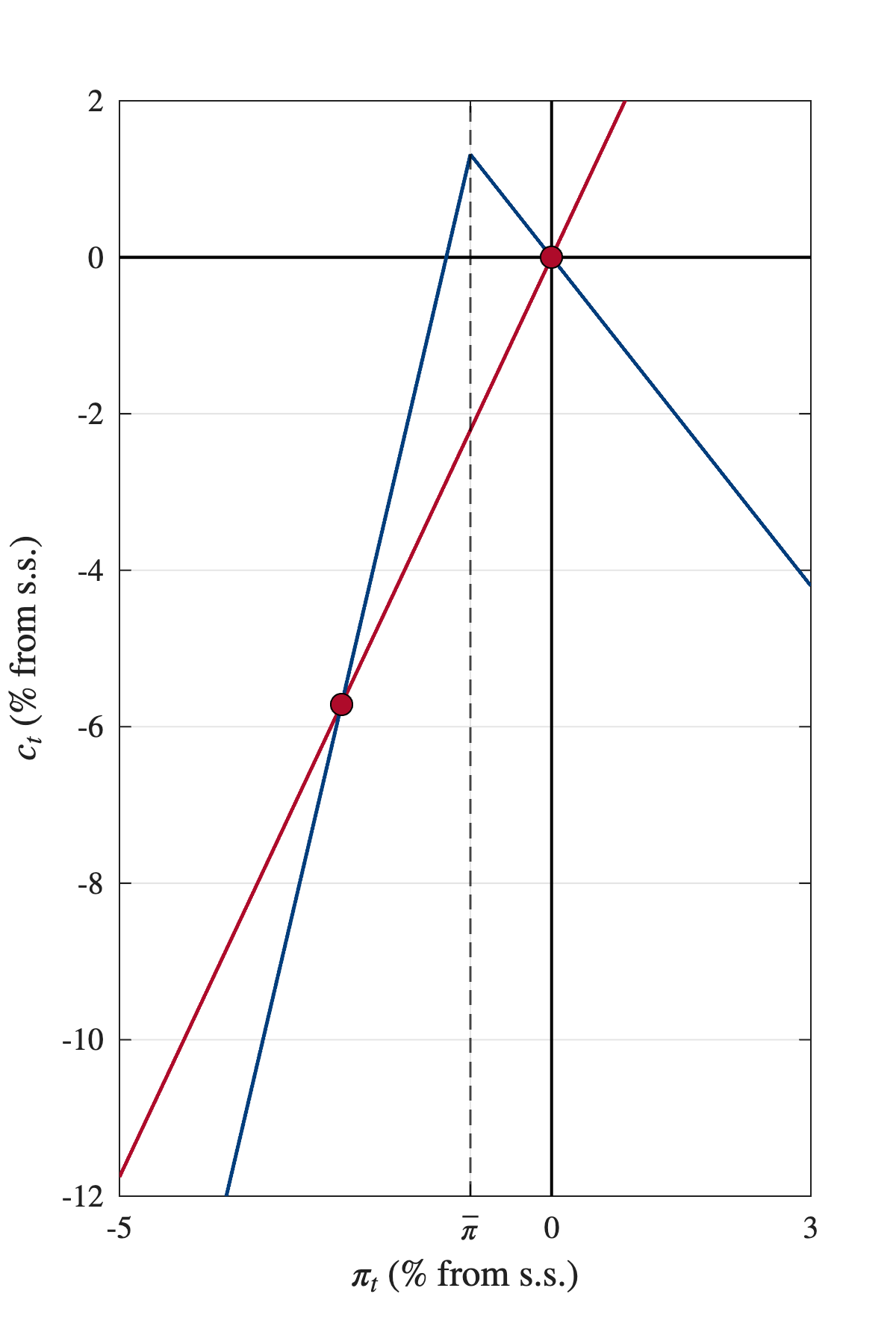}
    \caption{Case 2}
  \end{subfigure}\hfill
  \begin{subfigure}[t]{0.3\textwidth}
    \centering
    \includegraphics[width=\textwidth]{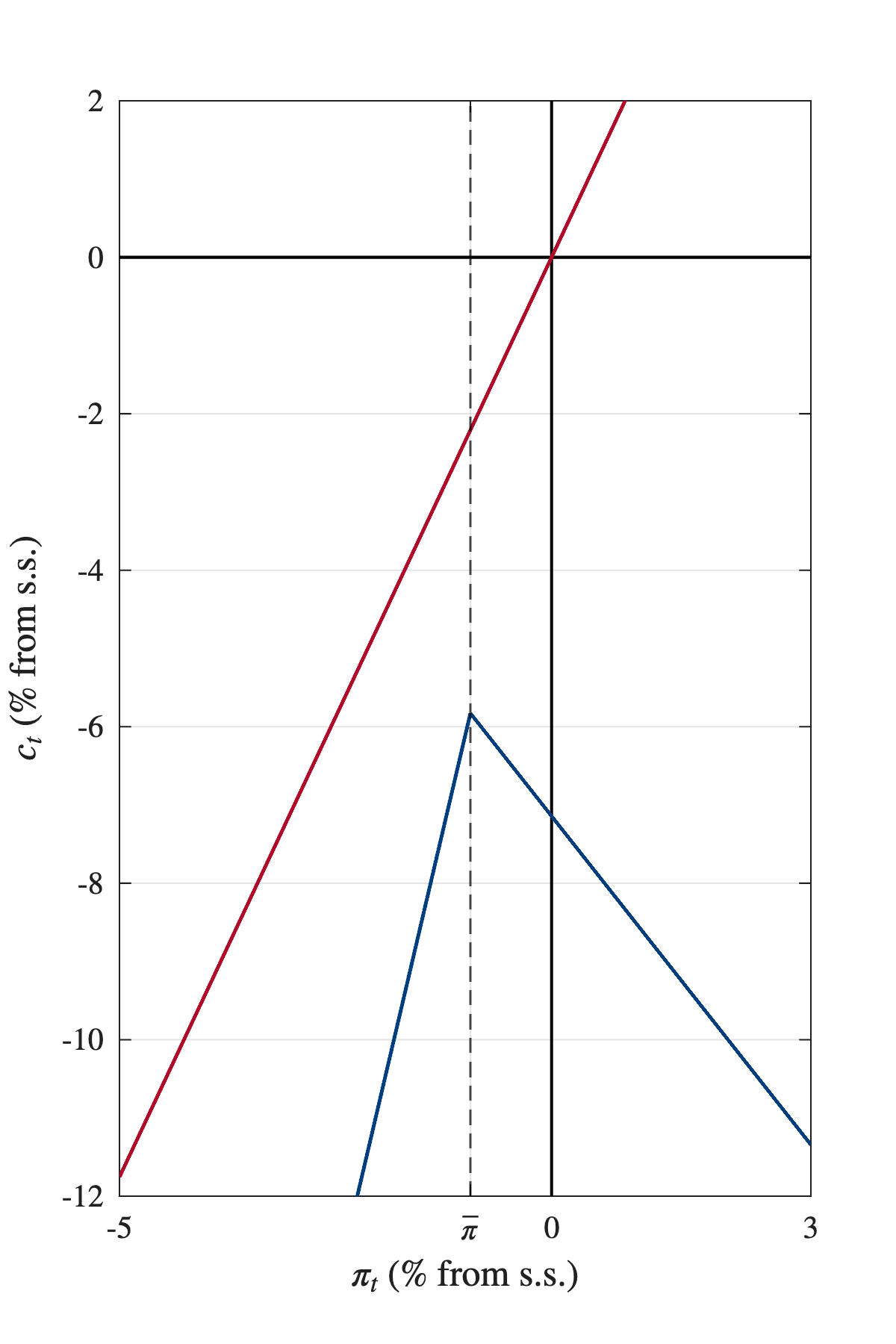}
    \caption{Case 3}
  \end{subfigure}
  \caption{AS and AD diagrams corresponding to the three cases in the literature.}
  \label{fig:3cases_AS_AD}
\end{figure}

These cases have been studied extensively in the literature. Case 1 is the standard one introduced in \cite{eggertsson2011fiscal} and that has received the most attention, see \cite{christiano2011government} and \cite{woodford2011simple} for two prominent examples. Case 2 is the one introduced in \cite{mertens2014fiscal} and that has also received a fair bit of attention, especially to explain the long liquidity traps in Japan. Following the terminology in \cite{Ascari2022unbearable} and \cite{ascari2023coherence}, we say that the model exhibits \textit{incompleteness}. Case 3 hasn't received nearly the same degree of attention. Following  \cite{Ascari2022unbearable} and \cite{ascari2023coherence}, we say that in this case the model exhibits \textit{incoherence}. In order to gain some intuition regarding what is happening in all those three cases, we illustrate the underlying aggregate supply/demand curves that give rise to each case in Figure \ref{fig:3cases_AS_AD}, where we express consumption as a function of inflation.
In all three panels, there is a threshold level of inflation $\overline{\pi}$ below which the Central Bank will have no choice but to set the nominal rate to 0. When that happens, the aggregate demand (AD) curve slopes positively because as the nominal rate doesn't move, the real interest rate moves one for one with the inflation rate: the Central Bank cannot accommodate variations in inflation. 
In the first panel, the negative demand shock generates lower inflation and private consumption by shifting the aggregate demand curve down. If this shift is large enough, then the two curves cross on the positive sloping part. It can be easily seen that only one MSV equilibrium clears the "verify" part in the guess-and-verify exercise. 

In the second panel, noting again that we express consumption as a function of inflation, the increased persistence of the underlying shock makes both slopes of aggregate demand larger in magnitude \textemdash remember that $\E_t c_{t+1} = p\cdot c_t$ in that framework. In that case there are two MSV equilibria that clear the "verify" part in the guess-and-verify exercise. \cite{mertens2014fiscal} then assume that a sunspot makes the agents coordinate on the one in which the ELB is binding. Following the existing literature, we assume that the sunspot is perfectly correlated with the demand shock. If that demand shock is small enough in magnitude, the shift down in AD is small enough to preserve the two crossings. As a result, the ELB equilibrium can be a result of the demand shock as well.

In the third panel, we depict what happens if the shift in aggregate demand is too large in magnitude. In that case, there are no MSV equilibria that clear the "verify" part in the guess-and-verify exercise. This case has been studied in detail in \cite{Ascari2022unbearable}. Potential fixes to avoid that situation from happening can be found in \cite{ascari2023coherence} as well as \cite{Murakami2026restoring}. The solution that we propose in this paper is largely complementary to these efforts. The main intuition that we will leverage is that these papers guess and verify whether there exists a \textit{constant} allocation that is perfectly correlated with the Markov Chain. Using a truncated Markov chain, we show that even though a constant allocation may fail to exist/be unique, a unique time-varying one is guaranteed to exist. 

To sum up the morass in one graph, we now provide a figure that represents which case arises for each value in $[p\ \ d]$ in $[0,1]\times[0,d^{max}]$ for a standard calibration of the model.\footnote{The calibration is given by $\sigma = 1$, $\beta = 0.99$, $\kappa = 0.05$, $\eta = 2$, $\psi = 2$, $\bar{c} = 0.8$, $m_{xx} = 0.95$, $m_{x \pi} = 0.95$, and $m_{\pi \pi} = 0.91$.} This figure shows that the counterintuitive behavior of the NK model with standard shocks happens to the right of $\overline{p}$.\footnote{The calibration that we use implies $\pbar=0.75$, which translates into an expected duration of exactly 1 year for the shock.} As a result, virtually all of the efforts to rid the model of its puzzles boil down to finding modifications of the model that push $\overline{p}$ \textit{to the right}. If one manages to push it above one, then the model can be guaranteed to produce a unique MSV solution. We will depart from this literature in two ways.

\begin{figure}[!htb]
  \centering
  \includegraphics[width=0.7\textwidth]{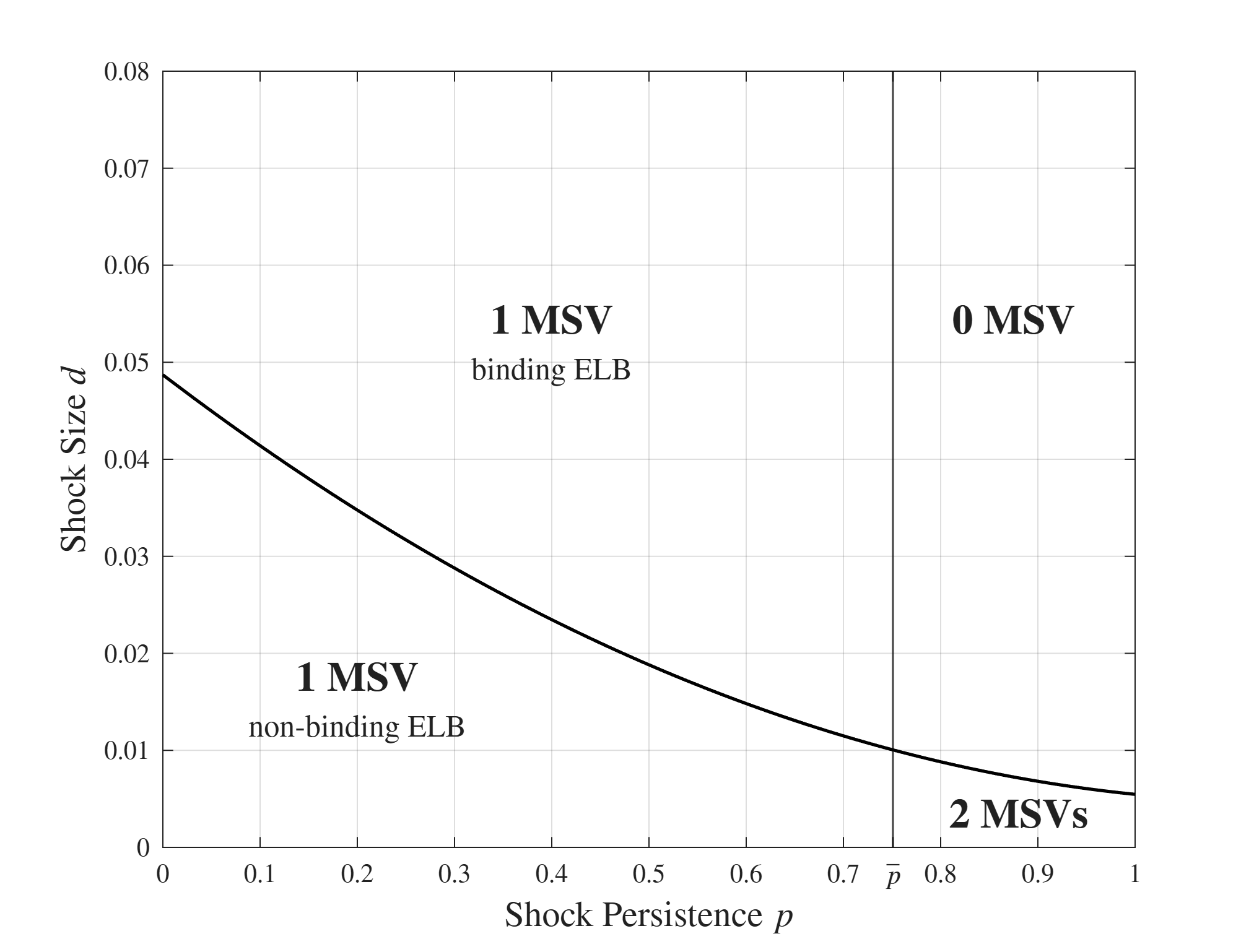}
  \caption{Regions of MSV Solutions in $(p, d)$ space.}
\label{fig:Regions}
\end{figure}

First, we will show that truncating the Markov chain results in a time-varying allocation that converges to a constant allocation as $\ell\to\infty$ only under some precise circumstances that we clarify. Second, we show that, in our framework with a truncated Markov chain, what is happening to the right of $\overline{p}$ at the ELB isn't different in sign, but only in \textit{magnitude}. 

Before moving on, note that the solution for the morass that we propose in this paper is specific to a setup with a standard New Keynesian model and one-off Markov chain shocks. As a result, this solution doesn't rule out the presence of confidence-driven liquidity traps in more elaborate New Keynesian models that may take the form of a "stagnation trap" as in \cite{Benigno2018stagnation} for example. In addition, given that we restrict ourselves to absorbing Markov chains, our results do not apply to a setting in which sunspots can be recurring as in \cite{Aruoba2018} or \cite{Nakata2022expectations}, who consider stochastic processes where the ELB episode can be recurring. While these are best suited to study the dynamics of inflation/macro aggregates, our formulation is more geared towards studying what happens after a one-off shock that mimics a recession.

\section{Resolving the Morass}
\label{sec:resolving_morass}

In this section we depart from most of the recent literature in that we follow \cite{eggertsson2003optimal} as well as \cite{eggertsson2021toolkit} and consider a truncated Markov chain.

\begin{definition}[Truncated Markov Chain]
We assume a stochastic process for $d_t$ with an initial value of $d$. For time periods $t,t+1,\dots,t+\ell-1$, the process follows a Markov chain with the transition matrix $\mathbf{P}$ introduced earlier. If the process is not back to its absorbing state of 0 after these $\ell$ periods, then it deterministically jumps to its absorbing state.
\end{definition}
Both \cite{eggertsson2003optimal} and \cite{eggertsson2021toolkit} use a large truncation to approximate a \textit{bona fide} absorbing Markov chain. They use this to solve numerically a general class of models that may include endogenous propagation mechanisms. Our focus is different in that we use this truncation to study explicitly the stability of equilibrium sequences in a standard New Keynesian model without endogenous propagation mechanisms. The truncation will also be useful in order to prove that the equilibrium sequence is in fact unique and that the multiplicity found in the existing literature is due to an analytical continuation as stated in the introduction. Finally, we also show that this truncation enables us to construct an AR(2) recursion for policy multipliers that further exemplifies why policy multipliers can diverge as a function of $\ell$ and do not converge to the multipliers found in the existing confidence-driven liquidity trap literature in the limit.

Formally, it can be shown that our specification nests as special cases two formulations that have been used in the existing literature. These are described in the following two corollaries.
\begin{corollary}
\label{cor:deterministic}
If $p=1$, then the shock becomes a deterministic one with an exact duration of $\ell$ time periods.    
\end{corollary}
Such a specification has been used in the context of continuous-time models in \cite{Werning2012managing} and \cite{Cochrane2017new}. It follows that the results that we establish later apply to the case of a deterministic shock.
\begin{corollary}
\label{cor:2state_Markov}
In the limit as $\ell\to\infty$, the stochastic process boils down to the standard 2-state absorbing Markov chain used in the literature.    
\end{corollary}

\noindent\textbf{Remark}\textemdash As stated in the introduction, what we call a "standard" absorbing Markov chain is one that essentially mimics a MIT shock. In that context, a standard Markov chain (without the "absorbing" adjective) would be one that has recurring states and that could potentially approximate an $AR(1)$ process following the methods developed in \cite{Tauchen1986finite}, \cite{Rouwenhorst1995asset} or \cite{Kopecky2010finite}.\footnote{Note that, as is shown in \cite{roulleau2023analyzing}, a carefully specified absorbing Markov chain such as the one in the current paper can be shown to exactly replicate, in expectations, the \textit{impulse response} of an arbitrary $AR(p)$ process. Hence the connection with MIT shock literature.} 

The main difference between the truncated Markov chain and the standard absorbing Markov chain used in the literature is that the latter lacks a terminal condition. Indeed, even though this happens with a vanishingly low probability, it is possible that a standard absorbing Markov chain remains in its low state for an arbitrarily long period of time. Given the forward-looking nature of the standard NK model, papers in the literature have to consider this case when computing conditional expectations. To say it differently, a Markov chain has a memory-less property. This means that, no matter how long the chain has been stuck in the low state, the probability to transition to the absorbing state is always $1-p$.

In contrast, under Assumption \ref{ass:intended_SS}, our truncated Markov chain features a well-defined terminal condition. That terminal condition allows us to use a standard backward induction method. To see why such a terminal condition is crucial, consider our Phillips curve equation \eqref{eq:NKPC}. Let us denote $\tilde{\pi}_t$ and $\tilde{c}_t$ as inflation and private consumption, conditional on the demand shock being in its low state. Notice that, assuming $\ell > 1$ we can rewrite it as follows:
\begin{align*}
\tilde{\pi}_t &= \kappa (1+\eta \cbar) \tilde{c}_t + \beta m_{\pi\pi}  \mathbb{E}_t \pi_{t+1}\\
&=  \kappa (1+\eta \cbar) \tilde{c}_t + \beta m_{\pi\pi} \left[p\cdot \tilde{\pi}_{t+1}+(1-p)\cdot 0\right],
\end{align*}
where we have used the fact that inflation is 0 in the absence of shocks in the intended steady state. For the sake of the argument, assume that the ELB is not binding at time $t$ so that we can write the nominal interest rate as a function of the inflation rate using the Taylor rule. Repeating the same procedure with the Euler equation, we end up with a system of two equations in \textit{four} unknowns: $\tilde{\pi}_t,\tilde{c}_t,\tilde{\pi}_{t+1}$ and $\tilde{c}_{t+1}$. In turn, both $\tilde{\pi}_{t+1}$ and $\tilde{c}_{t+1}$ depend on $\tilde{\pi}_{t+2}$ and $\tilde{c}_{t+2}$ by the same procedure. Because a standard absorbing Markov chain can stay in the low state for an arbitrarily long period of time, this process carries on \textit{ad infinitum}. 

In our setup, this doesn't happen. Assume that the Markov chain has remained in its low state for the $\ell$ periods $t,t+1,\dots,t+\ell-1$ and that $\ell \gg 1$. It follows that it will be forced to go back to its absorbing state next period. Therefore, we can write $\E_{t+\ell-1}\pi_{t+\ell}=\E_{t+\ell-1}c_{t+\ell}=0$, where $\E_{t+\ell-1}$ denotes expectations conditional on all information up to time period $t+\ell-1$. In that case, the Euler equation and Phillips curve become:
\begin{align}
\label{eq:Euler_ellm1}
\tilde{c}_{t+\ell-1} &=  - \sigma\cdot (\max \{\psi \tilde{\pi}_{t+\ell-1}, -\mu \} + d) \\
\tilde{\pi}_{t+\ell-1} &=  \kappa(1+\eta \cbar) \tilde{c}_{t+\ell-1}.
\label{eq:NKPC_ellm1}
\end{align}
Because of the max operator in the equation \eqref{eq:Euler_ellm1}, this is a non-linear system of two equations in two unknowns. Using the fact that equation \eqref{eq:Euler_ellm1} is piecewise linear however, we can show that this system of non-linear equations has a unique solution. We establish this now:

\begin{prop}
\label{prop:unique_ellm1}
The system of equations \eqref{eq:Euler_ellm1}-\eqref{eq:NKPC_ellm1} has a unique solution for a given value of $d$.    
\end{prop}
\begin{proof}
See online Appendix, section B. 
\end{proof}

Intuitively, the proof relies on the fact that one can transform this system of equations into a single piecewise linear equation in $\tilde{c}_{t+\ell-1}$. Then, notice that equation \eqref{eq:NKPC_ellm1} describes an upward sloping linear function crossing with either a horizontal line or a decreasing linear function. Perhaps more intuitively and in order to tie it back to the aggregate supply/demand graphs that can be found in \cite{eggertsson2011fiscal}, it is instructive to plot equations \eqref{eq:Euler_ellm1}-\eqref{eq:NKPC_ellm1} separately with $\tilde{\pi}_{t+\ell-1}$ on the $x$ axis.
\begin{figure}[!htb]
  \centering
  \begin{subfigure}[t]{0.48\textwidth}
    \centering
    \includegraphics[width=\textwidth]{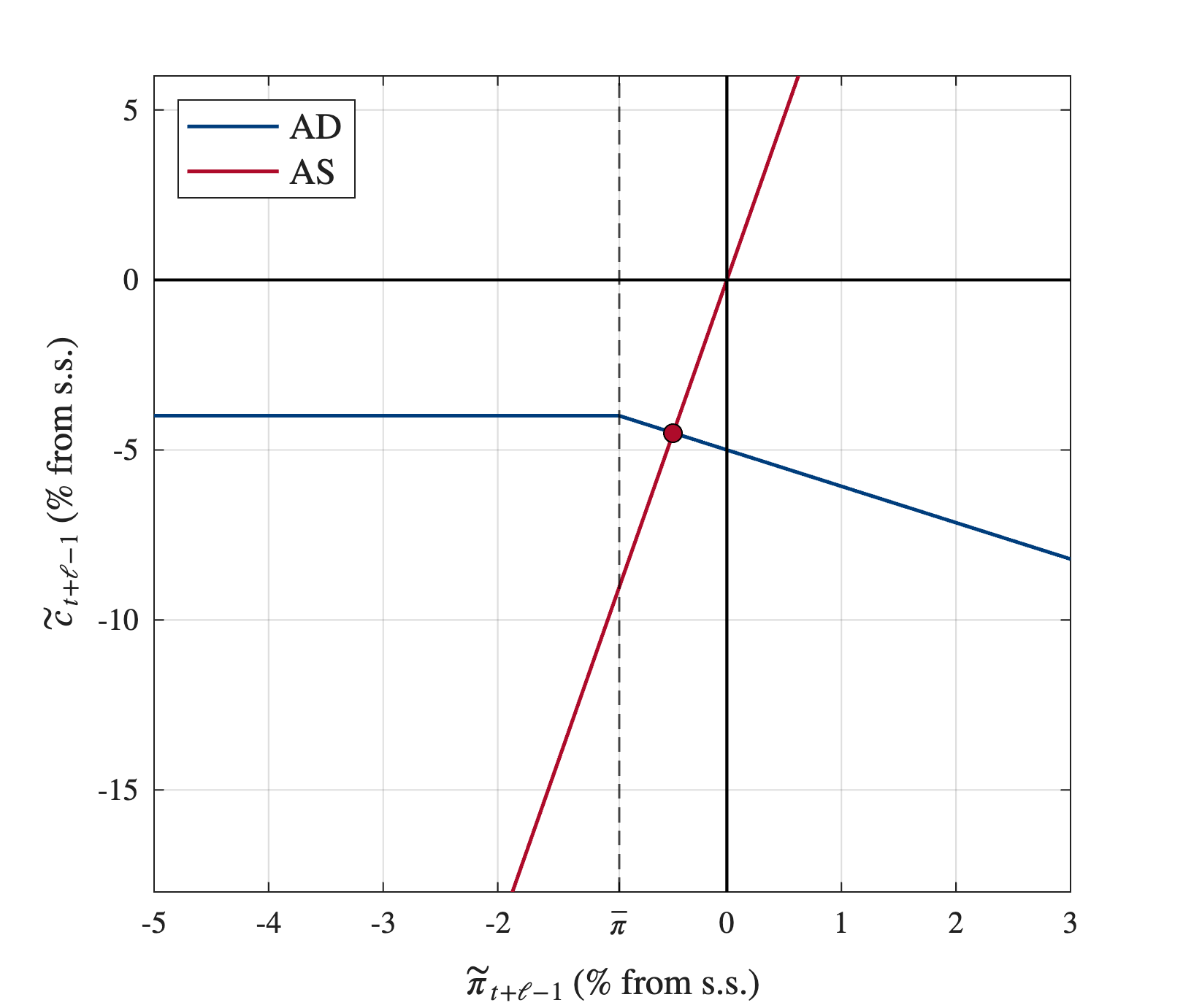}
    \caption{$d < \bar{d}(0)$}
  \end{subfigure}\hfill
  \begin{subfigure}[t]{0.48\textwidth}
    \centering
    \includegraphics[width=\textwidth]{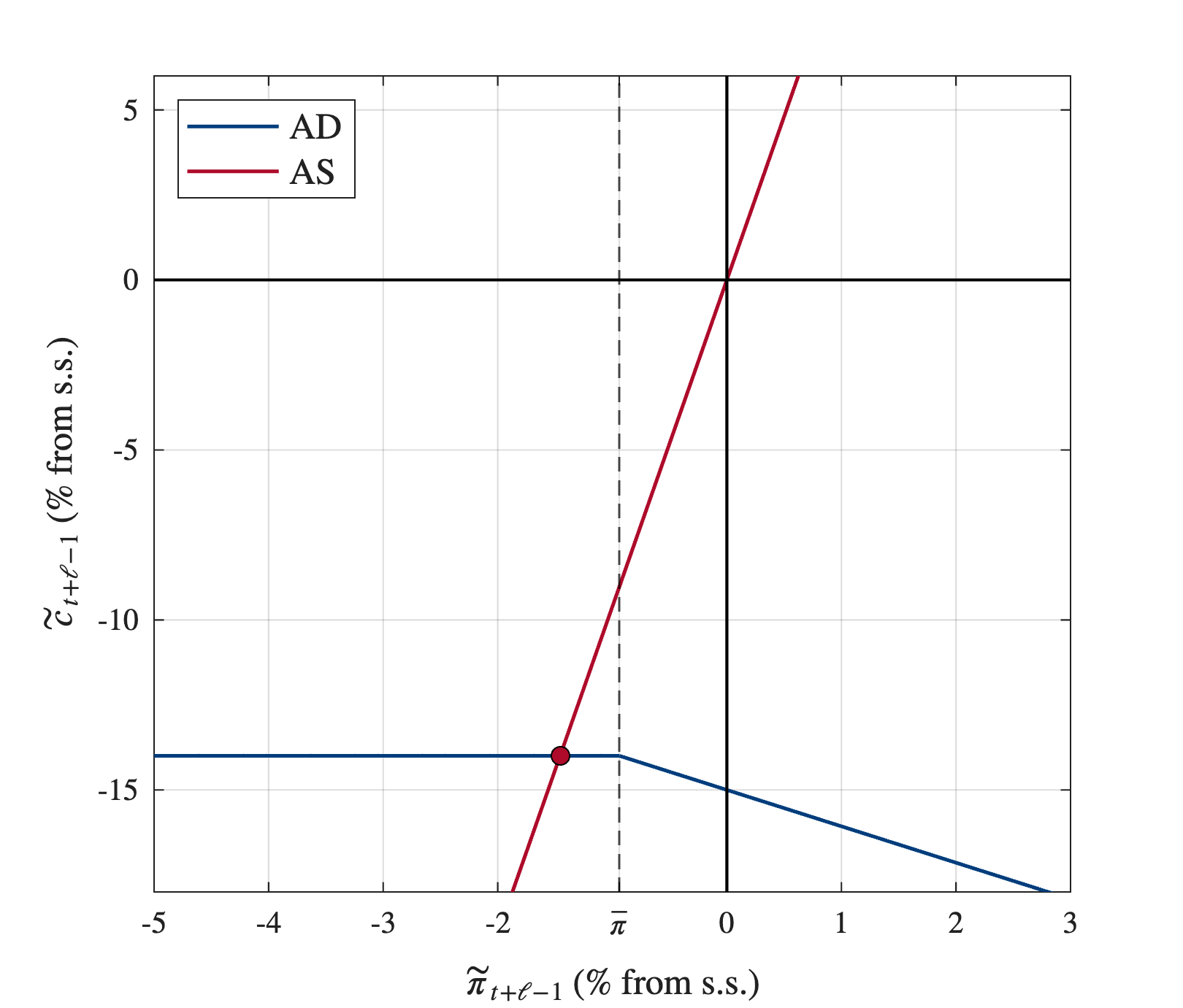}
    \caption{$d > \bar{d}(0)$}
  \end{subfigure}
  \caption{Aggregate Supply and Demand in Period $t+\ell-1$}
  \label{fig:tplm1_AS_AD}
\end{figure}%
The main take-away from Figure \ref{fig:tplm1_AS_AD} is that, because of the terminal condition, expectations at time $t+\ell-1$ are effectively exogenous.\footnote{This is where our setup markedly departs from the one in \cite{woodford2019monetary}, in which expectations right at the end of the planning horizon switch to being backward-looking as a function of all the realized history of endogenous variables. The exogeneity of expectations at time $t+\ell-1$ is what buys us the tractability that we leverage to characterize dynamics in closed form later on.} This is the reason why, in sharp contrast with the existing literature, the slope of AD at the ELB (to the left of $\overline{\pi}$) is zero. In our case, there is no automatic feedback between current realizations and expectations that can bend the AD curve back towards the aggregate supply curve at the ELB. Given that a realization of $d$ will shift the AD curve up or down, there can be only one crossing. In the existing literature, the standard absorbing Markov chain assumption means that, regardless of how long the shock has remained in its low state, it is still expected to be in the same state tomorrow with probability $p$.

Importantly, this intuition carries over to all time periods all the way back to the first one where the shock first materializes. To see this, note that both $\tilde{c}_{t+\ell-1}$ and $\tilde{\pi}_{t+\ell-1}$ are linear combinations of exogenous expectations upon exit and the shock $d$: as of time period $t+\ell-2$, both are effectively exogenous. As a result, the graphical representation of the allocation at time $t+\ell-2$ will be similar to the one in Figure 3, except that conditional expectations will now be non-zero. To see this, note that we can write the conditional expectation of inflation as $\E_{t+\ell-2}\pi_{t+\ell-1} = p\cdot\tilde{\pi}_{t+\ell-1}+(1-p)\cdot 0$,
which is again effectively exogenous as of time period $t+\ell-2$. The same goes for time periods $t+\ell-3,\dots,t$ assuming again that $\ell\gg1$. Using the fact that expectations are effectively exogenous considerably simplifies the backward induction process. In turn, it allows us to compute easily the whole path of inflation/private consumption for any given value of $\ell$. Ideally, one would want to know whether this path is unique or not. This is done in the next proposition, which is a generalization of proposition \ref{prop:unique_ellm1}.
\begin{prop}
\label{prop:unique_path}
For each value of $d\in [0,d^{max}]$ and a maximum horizon $\ell$, there exists a unique equilibrium sequence $\left\{c_{t+k},\pi_{t+k}\right\}_{k\geq 0}$ that solves equations \eqref{eq:Euler}, \eqref{eq:NKPC} and \eqref{eq:Taylor}.
\end{prop}
\begin{proof}
See online Appendix, section C.    
\end{proof}
Following the terminology laid out in \cite{Ascari2022unbearable} and \cite{ascari2023coherence}, Proposition \ref{prop:unique_path} guarantees that, under a truncated Markov chain, there is neither incoherence nor incompleteness. This finding is related to \cite{eggertsson2011fiscal}, who shows that $p<\pbar$ is a necessary and sufficient condition for a unique rational expectations MSV equilibrium. In this context, there can be sunspot equilibria when $p>\pbar$, but Proposition \ref{prop:unique_path} guarantees that there is a unique equilibrium sequence with a finite maximum horizon $\ell<\infty$ while there can be 0 or 2 when $\ell=+\infty$.\footnote{Technically speaking, the MSV criterion is a selection device to pick a particular equilibrium when there are several potential candidates. In both \cite{eggertsson2011fiscal} and \cite{mertens2014fiscal} the Taylor principle may not hold in which case there is an infinity of potential equilibria. In our case, given a terminal condition there is only one candidate so we do not need any selection device. Accordingly, we drop the MSV adjective.} In addition, this uniqueness result is conceptually related to the one obtained in \cite{Holden2023existence} in the sense that we share the backward induction from a terminal condition. However, \cite{Holden2023existence} studies a New Keynesian model with perfect foresight and, to the best of our knowledge, doesn't provide the stability conditions that we emphasize in this paper. With this result in mind, we now set out to characterize these paths. 

In order to do so, it will be useful to rewrite the model given by equations \eqref{eq:Euler}-\eqref{eq:NKPC} under the assumption that it stays outside the ELB the whole time in matrix form as follows $\tilde{Y}_{t} = p\cdot \mathbf{A}\tilde{Y}_{t+1} + C\cdot d$. To ensure that this model is well-behaved, we need to guarantee that the eigenvalues of matrix $p\cdot \mathbf{A}$ decay exponentially. This requires the following assumption:
\begin{assumption}
\label{asp:pA_decay}
The parameters in equations \eqref{eq:Euler}-\eqref{eq:NKPC} are such that
\begin{align*}
\psi > m_{x\pi} + \frac{(1 - m_{xx})(\beta m_{\pi\pi} - 1)}{\kappa(1+\eta\cbar) \sigma}    
\end{align*}
\end{assumption}
Assumption \ref{asp:pA_decay} guarantees that the eigenvalues of matrix $p\cdot \mathbf{A}$ have a modulus that is strictly less than 1. It can also be shown that assumption \ref{asp:pA_decay} actually coincides with the Taylor principle. With this in mind, the following definition will be useful to characterize the paths back to steady state:


\begin{definition}
\label{def:3paths}
For the standard New Keynesian model given by equations \eqref{eq:Euler}-\eqref{eq:Taylor} and the truncated Markov chain process for $d_t$, we focus on three types\footnote{There are possible paths in which the economy goes in and out of the ELB multiple times, but these are knife-edge cases. See the discussion surrounding Figure 4.} of possible solution paths: 
\begin{itemize}
    \item If $\tilde{\pi}_{t+n}(\ell)>\overline{\pi}$ for all $n=0,\dots,\ell-1$ we call it a \textbf{Normal-Time solution (N)}
    \item If $\tilde{\pi}_{t+n}(\ell)\leq\overline{\pi}$ for all $n=0,\dots,\ell-1$ we call it a \textbf{ELB solution (L)}
    \item If there exists a value $k$ such that $\tilde{\pi}_{t+n}(\ell)\leq\overline{\pi}$ for all $n=0,\dots,k$ and $\tilde{\pi}_{t+n}(\ell)>\overline{\pi}$ for all $n=k+1,\dots,\ell-1$ we call it a \textbf{Switching solution (S)}
    \end{itemize}
\end{definition}
Note that after a realization $d_t$, the effect on inflation will depend on the maximum duration $\ell$. As a result, we have made this dependence explicit in Definition \ref{def:3paths}. That dependence is also useful to precisely define what we mean by "puzzles" in the context of this paper. This is a loaded term that has taken on different meanings in the existing literature. For example,  \cite{Garin2019supply} refer to the fact that a decrease in productivity is expansionary at the ELB a puzzle. Our definition is closer in spirit to the literature on the "Forward Guidance Puzzle" (see \cite{Del2023forward}), where the impact effect on inflation/output grows arbitrarily large as the announcement is about a period further in time.
\begin{definition}
\label{def:puzzle}
We say that the model exhibits a puzzle if $\lim_{\ell\to\infty}|\tilde{\pi}_t(\ell)|=\infty.$
\end{definition}
In words, as the maximum possible duration of the shock grows, the impact effect on inflation grows unbounded. If $\ell$ is understood as a planned increase/decrease in the interest rate in the future, then Definition \ref{def:puzzle} is the Forward Guidance Puzzle. Note that this is not only about the effects of policy and can be a result of the demand shock itself. In this sense, it is also close in spirit to the way a puzzle is defined in \cite{Michaillat2021resolving}. With these definitions and our assumptions in hand, we first illustrate our main results graphically and then derive the relevant underlying propositions. 

\begin{center}
\begin{figure}[!htb]
  \centering
  \includegraphics[width=0.7\textwidth]{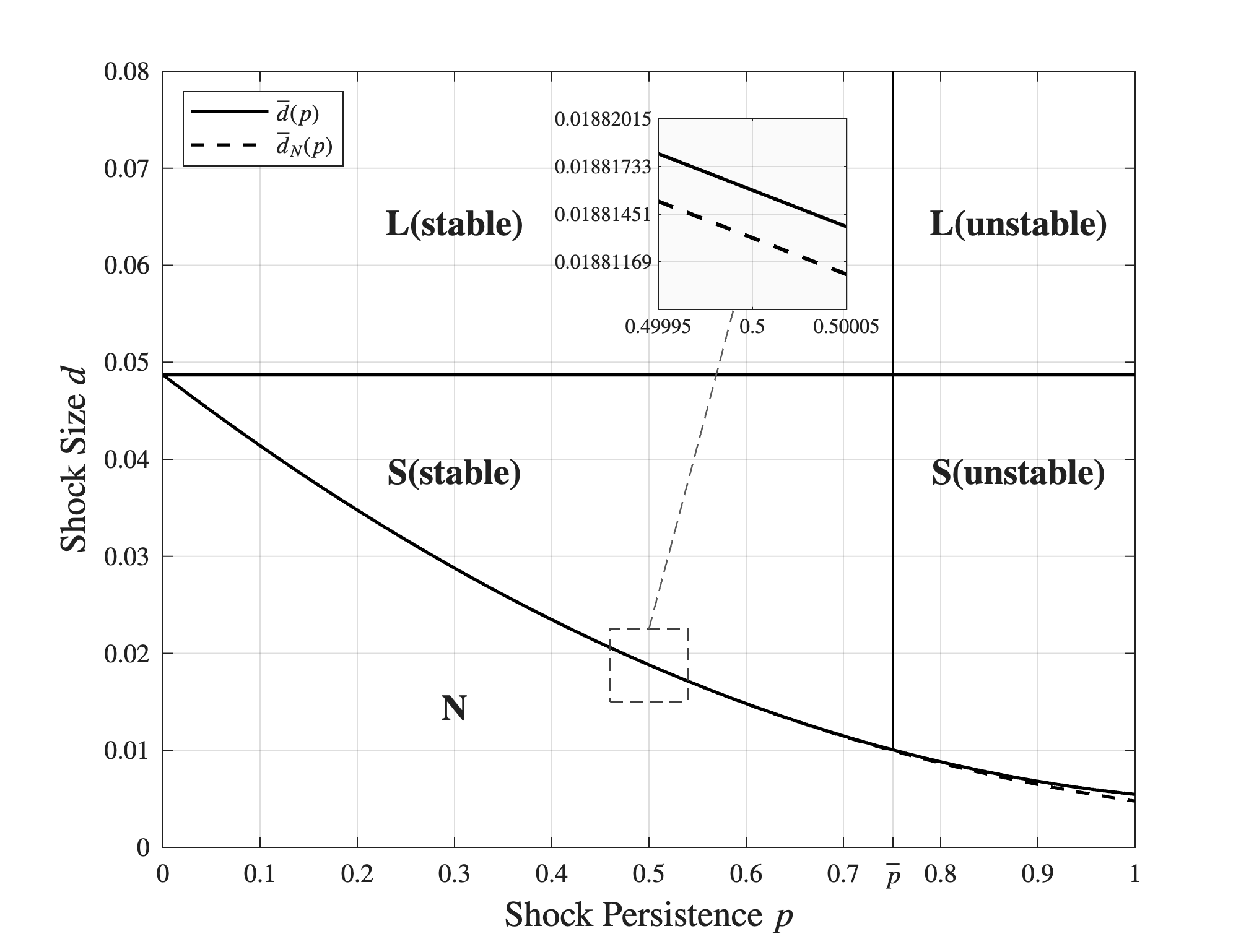}
  \caption{Types of Solutions in $(p, d)$ space for $\ell<\infty$.}
  \label{fig:Types}
\end{figure}
\end{center}

In order to visualize the possibilities listed in Definition \ref{def:3paths}, we provide a figure that is very close in spirit to Figure \ref{fig:Regions}. In sharp contrast with Figure \ref{fig:Regions} however, we do not flag the number of equilibrium solutions since there is only one for all parameter configurations. Instead, we flag the \textit{type} of the solution. Note that assumption \ref{asp:pA_decay} doesn't prevent the eigenvalues of matrix $p\cdot \mathbf{A}$ to be complex. When that happens, the standard New Keynesian model features a path for inflation that displays oscillations, possibly above and below $\overline{\pi}$. The immediate consequence is that the formula for $\overline{d}(p)$ derived in Proposition \ref{prop:existence_msv_equilibrium} doesn't exactly separate the cases where the solution is of $\mathbf{N}$ or $\mathbf{S}$ type. One can show however that, if $\overline{d}(p)<d<\overline{d}(0)$ and $\ell$ is sufficiently large, the equilibrium will be of type $\mathbf{S}$.  In this context, let $\overline{d}_N(p)$ denote the threshold value for the demand shock below which the economy stays outside the ELB for the whole path back to steady state. The presence of the aforementioned oscillations prevents us from computing this threshold analytically so we compute it numerically.\footnote{Under the assumption that the parameters in equations \eqref{eq:Euler}-\eqref{eq:NKPC} are such that $\psi < \frac{m_{x\pi}}{\beta m_{\pi\pi}}$, then it follows that $\overline{d}_N(p)=\overline{d}(p)$ and can thus be computed analytically. That is a knife-edge case however, because for the standard New Keynesian model this requires that $\psi<1/\beta$, which virtually all calibrations violate. In case that assumption doesn't hold, it can still be shown that $\overline{d}_N(p)\leq\overline{d}(p)$.} For clarity's sake, we display both the analytical and numerical thresholds in Figure \ref{fig:Types}. In our propositions later, we will leave out the cases where the duplet $(p,d)$ falls in between the two thresholds.\footnote{These will be cases where the solution can switch from binding to non-binding ELB multiple times along the transition back to steady state.} Note that these are virtually indistinguishable from each other, which implies that this omission is largely immaterial.

The main difference between Figures \ref{fig:Regions} and \ref{fig:Types} is that the first one assumes $\ell=\infty$ while the second one assumes $\ell<\infty$. Take the top-right region from Figure \ref{fig:Regions} where there is \textit{no} MSV equilibrium at all. This is because the researchers studying this configuration guess a \textit{constant} solution and then try to verify it. They find that no solution can verify that guess. Figure \ref{fig:Types} explains why that is the case: there exists a unique solution, but it is both unstable and time-varying. Even though the shock becomes a standard Markov chain as $\ell\to\infty$, the solution of the model does not converge to the guessed solution under a Markov chain.

Take now the region where there are two MSV equilibria whenever $\ell=+\infty$. In that case, researchers typically choose to focus on the one where the ELB is a binding constraint. There is however no compelling reason why this equilibrium should materialize instead of the one where the ELB doesn't bind. Our equilibrium construction resolves this issue entirely. Given a triplet $p,d,\ell$, one can construct the allocation right before the shock vanishes. If $d<\dbar_N(p)$, then our equilibrium construction picks the normal-time solution for all values of $\ell$. If $d > \dbar(p)$ instead, it picks a switching solution where the ELB binds on impact only if $\ell\geq \underline{\ell}$. Now that we've illustrated the different cases graphically, we turn to the relevant underlying propositions.


\begin{prop}[Non-switching solutions]
\label{prop:pure_solutions}
Assume that Assumptions~\ref{ass:intended_SS}
and~\ref{asp:pA_decay} hold. Then there exists a function
$\overline{d}_N
(p):(0,1)\to\mathbb{R}$ and a threshold
$\overline{p}\in(0,1)$ such that, for a given $p$,
\begin{itemize}
    \item if $d\in[0,\overline{d}_N(p))$, then the economy remains in the
    normal-time regime for all time periods, and
    $\lim_{\ell\to\infty}\tilde{\pi}_t(\ell)=\pi_t^N<\infty$ as well as
    $\lim_{\ell\to\infty}\tilde{c}_t(\ell)=c_t^N<\infty$;
    
    \item if $d\in[\overline{d}(0),d^{max}]$, then the economy remains in the
    ELB regime for all time periods;
    
    \item if $d\in[\overline{d}(0),d^{max}]$ and $p<\overline{p}$, then
    $\lim_{\ell\to\infty}\tilde{c}_t(\ell)=c_t^L$ and
    $\lim_{\ell\to\infty}\tilde{\pi}_t(\ell)=\pi_t^L$. If
    $d\in[\overline{d}(0),d^{max}]$ and $p\geq\overline{p}$, both variables
    diverge to $-\infty$ as $\ell\to\infty$.
\end{itemize}
where we have used superscripts $N$ and $L$ to denote the normal times and liquidity trap solutions constructed in \cite{eggertsson2011fiscal}. 
\end{prop}

\begin{proof}
See Appendix \ref{proof:pure_solutions}.
\end{proof}
The first part of Proposition \ref{prop:pure_solutions} states that, \textit{regardless of the value of $p$}, if the initial demand shock is not large enough in magnitude, then the ELB will never be a binding constraint. Intuitively speaking, this is the situation depicted in the left panel of Figure \ref{fig:tplm1_AS_AD}. Assuming that the shock is still alive before the maximum date, expectations of being in steady state next period combined with the shift down in AD due to the shock are not enough to drive that economy at the ELB: this is the unique guaranteed outcome from the guess and verify process. Further, the current decrease in inflation and private consumption will show up as expectations in the previous period. Once again however, these will not be sufficient to generate a binding ELB. It turns out that it holds for all previous periods. 

The second part of Proposition \ref{prop:pure_solutions} states that, once again \textit{regardless of the value of $p$}, if the shock is large enough in magnitude, then the ELB will bind in all periods. This is the situation depicted in the right panel of Figure \ref{fig:tplm1_AS_AD}. In contrast with the first part of the proposition, the shift down in AD is enough to make the ELB bind in the previous period. That intuition carries over to all previous periods. The third part of Proposition \ref{prop:pure_solutions} further categorizes the ELB solution into two distinct sub-regimes. In an ELB regime, the dynamics of the economy can be written as $\tilde{Y}_{t} = p\cdot \mathbf{A}^*\tilde{Y}_{t+1} + C^* \cdot d + E^*$ where the matrix $\mathbf{A}^*$ encodes the fact that the Taylor rule is now passive and $E^\ast$ collects the constant $\sigma\mu$ in the Euler equation. In the event that $p>\pbar$, at least one of the eigenvalues of $p\mathbf{A}^*$ is larger than 1 in modulus. When that happens, the backward induction algorithm generates an impact effect that is increasing in magnitude with $\ell$. In contrast, when $p<\pbar$ instead, the backward induction algorithm generates an impact effect that is finite and well-defined. In both cases, given a finite value of $\ell$, the equilibrium is finite and well-defined. 

Note that proposition \ref{prop:pure_solutions} is tightly linked to the familiar graphs depicted in Figure \ref{fig:3cases_AS_AD}. Indeed, $p>\pbar$ is equivalent to the path being potentially unstable \textit{and} the fact that the MSV equilibrium under a standard Markov chain is either non-existent or non-unique. Now that we have covered the non-switching solutions, we move on to the characterization of the switching solutions.

\begin{prop}[Switching solutions]
\label{prop:Mixed_solutions}
Assume that Assumptions~\ref{ass:intended_SS}
and~\ref{asp:pA_decay} hold. Then there exist a strictly decreasing function
$\overline{d}(p):(0,1)\to\mathbb{R}$ and a threshold
$\overline{p}\in(0,1)$ such that, for a given value of $p$,
\begin{itemize}
    \item if $\overline{d}(p)<d<\dbar(0)$, then there exists a finite integer
    $\underline{\ell}$ such that the ELB binds on impact if
    $\ell\geq\underline{\ell}$. In this case, the equilibrium path contains a
    single switch from the ELB regime to the normal-time regime. If
    $\ell<\underline{\ell}$, the ELB never binds;
    
    \item if $\overline{d}(p)<d<\dbar(0)$ and $p<\pbar$, then
    $\lim_{\ell\to\infty}\tilde{\pi}_t(\ell)=\pi_t^L$ from Proposition
    \ref{prop:pure_solutions}, and likewise
    $\lim_{\ell\to\infty}\tilde{c}_t(\ell)=c_t^L$;
    
    \item if $\overline{d}(p)<d<\dbar(0)$ and $p\geq\pbar$, then
    $\lim_{\ell\to\infty}\tilde{\pi}_t(\ell)=-\infty$.
\end{itemize}
As before, superscripts $N$ and $L$ denote the normal-time and liquidity-trap solutions developed in \cite{eggertsson2011fiscal}.
\end{prop}


\begin{proof}
See Appendix \ref{proof:mixed_solutions}.
\end{proof}

The first part of Proposition \ref{prop:Mixed_solutions} establishes that, if the shock $d$ lies in the specified range then whether the ELB binds on impact or not depends on the value of $\ell$. For the sake of the argument, assume that the ELB doesn't bind for the period before the shock vanishes. Given that $d$ is a negative demand shock, it will result in lower inflation/consumption compared to the non-stochastic steady state. Now think about the period before that. The shock is there by definition, and now there is expected deflation/lower expected private consumption. As a result, inflation and the consumption are mechanically lower than the period right after. That goes for the period before, and so and so on. If $d<\dbar_N(p)$, then that backward induction chain results in $\pi_t^N$\textemdash see Proposition \ref{prop:pure_solutions}. If $\dbar(p)<d<\dbar(0)$, one can guess a hypothetical normal-time solution. However, in that case the backward induction converges to a value of $\pi_t$ that is lower than $\pibar$ and thus violates the guess of a normal-time solution.    

So we know that if $d$ is in that range, eventually the expected deflation in the future will percolate back by enough to send the economy at the ELB in the short run. If $p<\pbar$, then that process converges to the same value of $\pi_t^L$ from Proposition \ref{prop:pure_solutions} in the limit as $\ell\to\infty$. Intuitively, for a finite value of $\ell$, the ELB will be binding on impact but the ELB will stop binding before the demand shock reverts to its steady state value. As before, if $p\geq \pbar$, the backward induction process gives a sequence that is diverging to $-\infty$. 

\begin{figure}[!htb]
  \centering
  \begin{subfigure}[t]{0.48\textwidth}
    \centering
    \includegraphics[width=\textwidth]{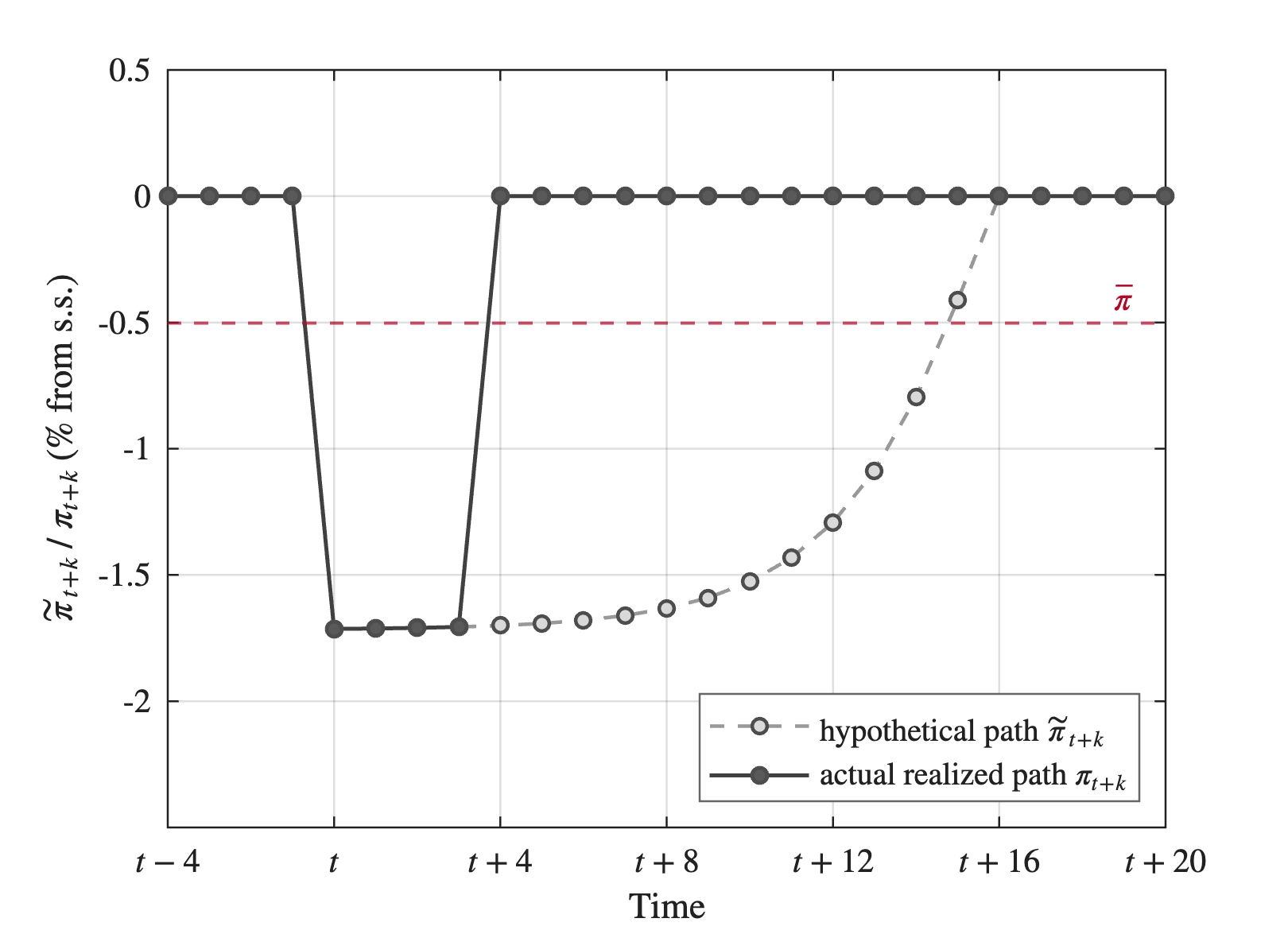}
    \caption{$p < \bar{p}$}
  \end{subfigure}\hfill
  \begin{subfigure}[t]{0.48\textwidth}
    \centering
    \includegraphics[width=\textwidth]{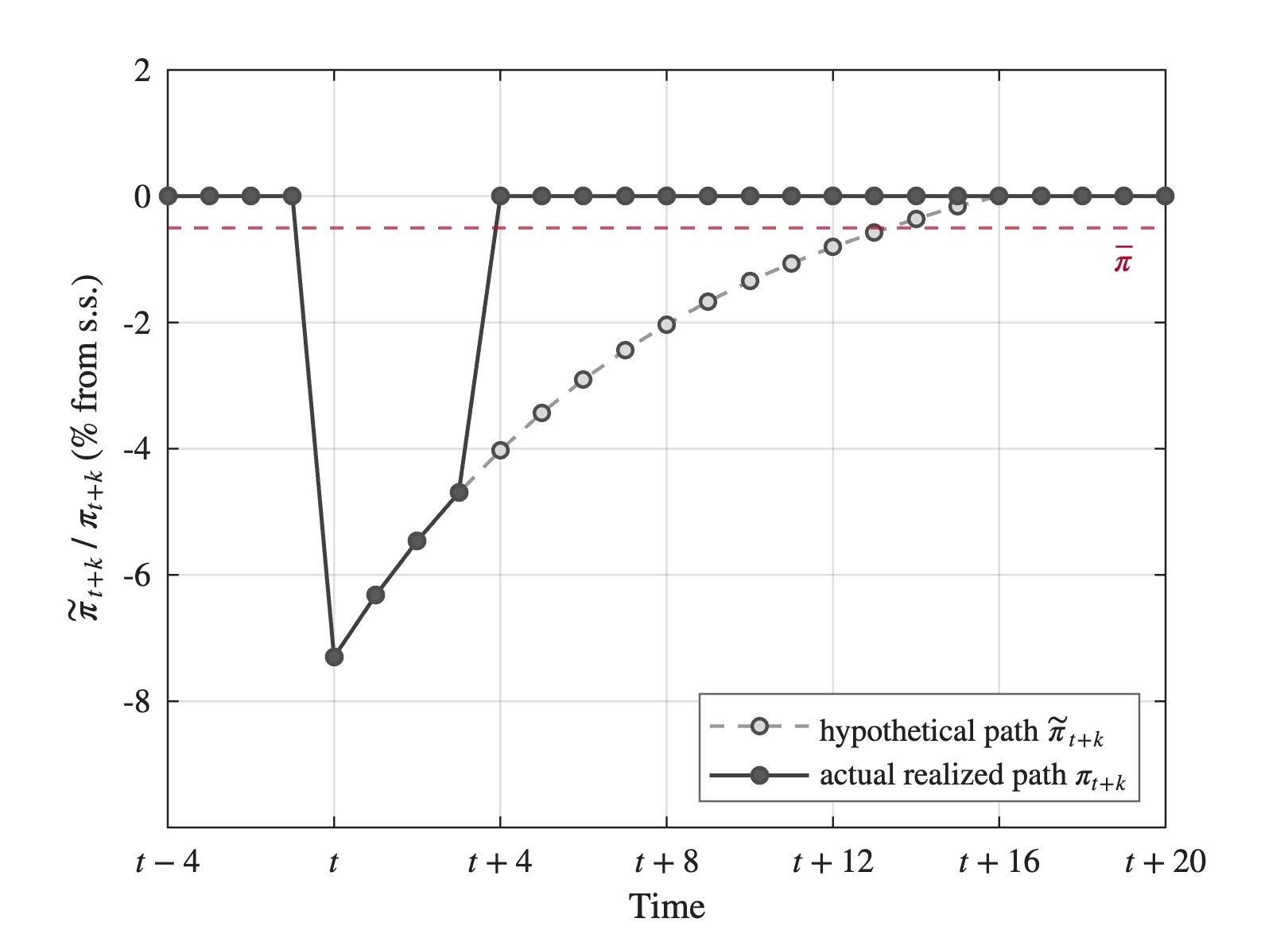}
    \caption{$p > \bar{p}$}
  \end{subfigure}
  \caption{Switching Solution: Realized Inflation Path for a Four-Quarter Shock with $\ell = 16$}
  \label{fig:two_paths}
\end{figure}

To illustrate these features, we report two experiments in Figure \ref{fig:two_paths}. We assume that  $\ell$ takes on a relatively large value of 16, which represents 4 years for a quarterly calibration. Given that this is the maximum number of periods during which the shock can be stuck in its low state, we assume that the shock actually reverts back at time $t+3$ after 4 quarters or 1 year. We consider two different cases, $p<\pbar$ (left panel) as well as $p>\pbar$ (right panel). In both cases, we report actual realized inflation $\pi_{t+k}(\ell)$ alongside the hypothetical path of inflation if the shock were to stay in its low state for the full $\ell$ periods.
Focusing on the left panel first, we note that for a low realized duration of just one year, the equilibrium path looks similar to the one we would get under a standard Markov chain: inflation looks like it is decreasing on impact to a constant level $\pi^L$ and stays there for 4 periods until shooting back up to its steady state of 0 as soon as the shock vanishes. Strictly speaking, this is not what is happening here. Instead, the equilibrium path for inflation is not constant over time. It is clearly apparent that, if the shock were to last for 3 years until time period $t+11$, the path of inflation would be inching its way up towards 0 and in fact would be above $\overline{\pi}$. In the existing literature which focuses on the limit as $\ell\to\infty$, that slow climb towards 0 never actually happens and what is left is the almost constant path that shows up for the first few periods in Figure \ref{fig:two_paths}. In the limit as $\ell=+\infty$, that path actually becomes constant. 

One can see immediately that this is not what is happening as soon as $p>\pbar$. In that case, the unstable dynamics are such that the economy never settles down to a constant equilibrium in the short or medium run. As $\ell$ grows larger, the impact effect on inflation grows more and more negative. For the following periods, inflation climbs back up and shoots back to steady state immediately if the shock subsides after less than $\ell$ periods. The papers in the existing literature do not recover this equilibrium path because they focus on constant allocations.

Taken together, propositions \ref{prop:pure_solutions} and \ref{prop:Mixed_solutions} characterize what can happen in this model in the sense that they uniquely characterize an equilibrium sequence for a given triplet $p,d,\ell$, except for the knife-edge cases in-between the thresholds $\overline{d}_N(p)$ and $\overline{d}(p)$. Note that Propositions \ref{prop:unique_ellm1}-\ref{prop:Mixed_solutions} are derived under the assumption that the economy goes back to the intended steady state. It turns out that this assumption, while natural, is not crucial for our results. Indeed, we show in the following proposition that even if the economy is assumed to go back to the unintended steady state then the equilibrium sequence for a given triplet $p,d,\ell$ is still unique.
\begin{prop}
\label{prop:unique_path_unintended}
Assume that the absorbing state is given by $\mathbf{Y}^u=\left[c^u\ \pi^u\right]$ where $c^u$ and $\pi^u$ denote consumption and inflation in the unintended steady state\textemdash expressed in deviations from the intended steady state. As a result, we have $\pi^u<\overline{\pi}<0$. Then, for each value of $d\in [0,d^{max}]$ and a maximum horizon $\ell$, there exists a unique equilibrium sequence $\left\{c_{t+k},\pi_{t+k}\right\}_{k\geq 0}$ that solves equations \eqref{eq:Euler}, \eqref{eq:NKPC} and \eqref{eq:Taylor}.

In addition, given this terminal condition there exists neither a normal time (type $\mathbf{N}$) nor a switching (type $\mathbf{S}$) solution, only the ELB (type $\mathbf{L}$) survives. For $d>0$, as before, the ELB solution is stable if $p<\pbar$ and unstable otherwise.
\end{prop}
\begin{proof}
See the online Appendix, section G.1.    
\end{proof}
It then follows that one can still construct a unique path that converges back to the unintended steady state in the end. In that case, the second part of the proposition guarantees that the ELB has to be binding both in the short and in the long run. This is due to the fact that the expectation to end up in the unintended steady state will bring about deflationary expectations in the periods before. This by itself is enough to guarantee that the economy is at the ELB the period before the return to the unintended steady state, which immediately rules out a solution of type $\mathbf{N}$. Furthermore, this effect compounds going back in time so that inflation decreases and the economy can never be outside the ELB in the short run. This in turn rules out solutions of type $\mathbf{S}$. If $p<\pbar$, inflation settles to a negative number as $\ell\to\infty$. If $p>\pbar$ instead, inflation diverges to $-\infty$.

\noindent \textbf{Discussion}\textemdash  
When there is a unique MSV equilibrium solution under a standard absorbing Markov chain, our equilibrium construction gives the same as $\ell\to\infty$. Therefore, discrepancies between our equilibrium construction and the one in the literature in the limit only arise if $p>\pbar$: the literature finds a stable, constant solution while we find a time-varying, unstable one. As alluded to in the introduction, the standard practice in the literature is to consider an \textit{analytical continuation}. More precisely, the impact effect of the shock on inflation can be shown to be given by the following expression:
\begin{align*}
\tilde{\pi}_t (\ell) &= V_2^{\top} \left[\sum_{i=0}^{\ell-1} (p\mathbf{A}^*)^{i} \right] (C^* \cdot d + E^*)
\end{align*}
where $V_2 = \left[0\ \ 1\right]^{\top}$ is a selection vector. The radius of convergence for this series is given by $\rho(p\mathbf{A}^\ast)<1$, where $\rho(\cdot)$ denotes the spectral radius (maximum absolute eigenvalue). As $\ell\to\infty$, if $\rho(p\mathbf{A}^\ast)<1$ then the matrix geometric series converges to
\begin{align}
\tilde{\pi}_t (\infty) &= V_2^{\top} (I - p\mathbf{A}^*)^{-1} (C^* \cdot d + E^*)
\label{eq:analytical_cont}
\end{align}
If $\rho(p\mathbf{A}^\ast)>1$, the matrix geometric series diverges instead. In that context the analytical continuation given by equation \eqref{eq:analytical_cont} is still well defined as long as $\det(p\mathbf{A}^\ast-I)\neq 0$. This is where we depart with the existing literature: we choose to keep $\ell$ finite and work with the partial geometric sum while in the existing literature the usual choice is to take $\ell=+\infty$ and work with the analytical continuation.

Relatedly, we think that our results should be interpreted as follows. In \cite{eggertsson2003optimal} and \cite{eggertsson2021toolkit}, a truncated Markov chain with a large $\ell$ is used to approximate a standard Markov chain. Instead, we argue that whenever $p<\pbar$ the standard Markov chain should be seen as an approximation of the short run dynamics when $\ell$ is moderately large. In that case,  the equilibrium in the short run will be approximately constant and can be summarized with a single constant number when $\ell=+\infty$. This shares a similarity with the fact that macroeconomists usually work with infinite horizon models compared to finite lifetime models: the first type of models give time-invariant policy functions that are easily characterized while the second type of models give time-varying ones that are more difficult to deal with. In the case where $p>\pbar$ that limiting argument doesn't work anymore and we argue that a relatively small duration for the shock enables the model to produce reasonable dynamics.

Beyond these discrepancies, all the situations that we have highlighted share the same qualitative property: both inflation and consumption \textit{fall} on impact. The magnitude may vary across cases, but the decrease is there no matter what. Given the findings reported in \cite{eggertsson2011fiscal} as well as \cite{mertens2014fiscal} and all the literature that ensued, the same clearly cannot be said about the effects of fiscal policy depending on whether $p\lessgtr\pbar$. This begs the following question: what does our equilibrium construction predict for the effects of government spending at the ELB? 

Another reason for focusing on the effects of government spending at the ELB comes from Propositions \ref{prop:pure_solutions} and \ref{prop:Mixed_solutions}, in which it is apparent that the potential for arbitrarily large outcomes in our model consists in the region $[\overline{p}\ \ 1]\times [\overline{d}(p)\ \ d^{max}]$. This is in sharp contrast with the existing literature in which the puzzles materialize when $p$ is in a neighborhood of $\overline{p}$. The natural question that arises is how big of a difference is there between the two regions? Given the extensive empirical literature on the effects of government spending at the ELB, we will use that as our metric: what is the region in the parameter space for which the model produces government spending multipliers outside of the credible empirical range?  In order to answer that question, we need to include government expenses back into the model. Accordingly, we jump to the effects of fiscal policy next.

\section{Characterizing Fiscal Policy at the ELB}
\label{sec:policies}

The framework that we have developed in the previous section gives a unique answer to the question: what happens after the economy is hit with a negative demand shock? In that setup, as we have seen, the economy ends up at the ELB if the shock is both large and persistent enough. We do not need to invoke a sunspot for that to happen. Therefore, this framework helps us understand \textit{how} and under which conditions the economy ends up at the ELB.

Ultimately however, policymakers want to know what to do once they find themselves in that position. In other words, one would want to understand the effect of policy in that context. In the existing literature, the standard New Keynesian model only gives mixed recommendations: if the economy is at the ELB because of a mildly persistent demand shock, then increasing government spending will crowd private consumption in. If the economy is at the ELB because there are two MSV equilibria and we pick the one with a binding ELB, then increasing government spending crowds private consumption out. The objective of this section is to show that the framework that we have developed in the previous section does \textit{not} give mixed recommendations at the ELB: the sign of the effect on consumption doesn't flip depending on how persistent the shock is.

Before we go on to describe the effects of government spending at the ELB, we first establish some results about the government spending multiplier in the normal-time regime. Given the result in the previous section, this regime applies if the demand shock is such that $d<\dbar_N(p)$ regardless of the value of $p$. In what follows we will focus our attention on the impact effect on private consumption. Indeed, the government spending multiplier effect on output will be strictly larger/smaller than 1 if and only if the effect on consumption is strictly larger/smaller than 0. In that context, it will be useful to define $m^{c,R}_{g}(\ell)$ as the \textit{impact} multiplier effect on consumption $c$ for a given $\ell$ and where the regime $R$ takes values in $\left\{N,L,S\right\}$. We characterize the multiplier as a function of $\ell$ now:


\begin{prop}[Multiplier in Normal Times]
\label{prop:mcg_cases_PN}
Assume that assumption \ref{asp:pA_decay} holds, that for a given $p$,
$d<\overline{d}_N(p)$ and that $\psi>pm_{x\pi}$. Then the impact effect on private consumption
$m^{c,N}_{g}(\ell)$ is such that $\lim_{\ell\to\infty}m^{c,N}_{g}(\ell)
    =m^{c,N}_{g}(\infty)<0$ and $m^{c,N}_{g}(\ell)<0$ for all finite $\ell\geq1$.
\end{prop}

\begin{proof}
Given our focus on the multiplier at the ELB, we relegate this proof to the online Appendix, section D.    
\end{proof}

The main take-away from Proposition \ref{prop:mcg_cases_PN} is that, regardless of the maximum duration $\ell$, the government spending multiplier on consumption is negative and it converges to the value $m^{c,N}_{g}<0$ that obtains using a standard Markov chain as $\ell\to\infty$. As a result, there is no scope for qualitative change in the sign of the multiplier as a function of $\ell$ outside the ELB as long as we maintain the restriction that $\psi>pm_{x\pi}$. This restriction is automatically satisfied in the standard New Keynesian model in which $m_{x\pi}=1$ and $\psi>1$. Note that for a behavioral model it is possible that the structural parameters are such that $\psi<pm_{x\pi}$. In that case, private consumption is crowded in outside the ELB but virtually all standard calibrations rule that feature out. Therefore, we argue that outside the ELB the typical outcome is that consumption is crowded out after an increase in government spending.

The existing literature has established that this may not be the case when one considers a demand shock that is large and persistent enough to send the economy in the ELB regime. For the sake of exposition, we begin by assuming that the demand shock is large enough such that the economy ends up in an ELB regime. Without imposing any restrictions on the absorbing state, this amounts to say that the demand shock is such that $d \geq \dbar(0,\pi^{ss})$ where we have made the dependence of the threshold on steady state inflation $\pi^{ss}$ explicit. Indeed, we have implicitly assumed that $\pi^{ss}=0$ throughout the paper. It can be shown however that $\dbar(0,0) > \dbar(0,\pi^u)$: in words, if the economy returns to the unintended steady state then it takes a demand shock of a lower magnitude to send the economy to the ELB in the short run. This is because the expectation to return to the unintended steady state once the demand shock is over adds a deflationary pressure in expectations. With this in mind, we show in the following proposition that conditional on being at the ELB consumption is crowded in, regardless of the steady state assumption. This assumption only matters insofar as it influences the possibility to be at the ELB in the first place for a given shock size.\footnote{As is standard in the literature, we assume that the increase in government spending is not big enough to get the economy out of the ELB.} Conditional on being at the ELB, the steady state assumption merely introduces a set of constants that disappear as we take the partial derivative with respect to $g$.

\begin{prop}[Multiplier in ELB]
\label{prop:mcg_cases_PL}
Assume that assumption \ref{asp:pA_decay} holds, that
$d > \dbar(0,\pi^{ss})$, and that $p\in(0,1]$. Then it follows that, regardless of $p$, the impact consumption multiplier $m^{c,L}_{g}(\ell)$ is positive and strictly increasing in $\ell$. In addition, for given initial conditions $m^{c,L}_{g}(1), m^{c,L}_{g}(2)$, the impact consumption multiplier can be written as the following AR(2) process\footnote{where the constant $c^*_m$ is defined by the time-invariant ELB multiplier $m_g^{c,L}(\infty)$ (i.e., the fixed point of the recursion, which coincides with the limit when $p < \pbar$) and the coefficients from the characteristic polynomial of $p \mathbf{A^*}$:
\[
    c^*_m \defeq \left(1 - \tau_p^* + \delta_p^*\right) m_g^{c,L}(\infty)
\]
In this expression, $\tau_p^* = \tr(p\mathbf{A}^*)$ is the trace of the system matrix $p\mathbf{A}^*$, and $\delta_p^* = \det(p\mathbf{A}^*)$ is its determinant.}:
\begin{align*}
m_g^{c,L}(\ell+2) = \tau_p^* m_g^{c,L}(\ell+1) - \delta_p^* m_g^{c,L}(\ell) + c^*_m
\end{align*}
for $\ell\geq 1$. Further, there exists a threshold $\pbar$ such that:
\begin{itemize}
    \item $\lim_{\ell\to\infty} m^{c,L}_{g}(\ell) = m^{c,L}_{g}(\infty) >0$ if $p<\pbar$
    \item $\lim_{\ell\to\infty}m^{c,L}_{g}(\ell)= +\infty$ if $p\geq\pbar$
\end{itemize}
\end{prop}
\begin{proof}
See online Appendix, section E.    
\end{proof}
The main take-away here is that the consumption multiplier is guaranteed to be positive at the ELB. This is a consequence of our earlier findings in the previous section. Another feature of Proposition \ref{prop:mcg_cases_PL} worth flagging is the AR(2) representation for the impact multiplier on consumption that we alluded to in the introduction. It turns out that the solution constructed in \cite{mertens2014fiscal} is a \textit{bona fide} fixed point of this recursion, but an unstable one. Indeed, the fact that $\rho(p\mathbf{A}^\ast)>1$ prevents the AR(2) from converging to that fixed point: it diverges instead. Using standard methods, one can solve for $m^{c,L}_{g}(\ell)$ and obtain a geometric sum. As before, the solution considered in \cite{mertens2014fiscal} is the analytical continuation defined outside the convergence radius of the geometric sum. This is the reason why their construction does not arise as an equilibrium in our setup with finite $\ell$ \textemdash even if that $\ell$ is made to be arbitrarily large. As a result, the result of a crowding-out effect highlighted in \cite{mertens2014fiscal} does not arise here. For the same parameter configuration that was employed in that paper, the effect on private consumption is both positive and growing with $\ell$. As a result, for once in Economics the answer is not "it depends": a policy maker using this model necessarily concludes that fiscal policy has a positive effect on private consumption at the ELB. 

The findings that we have reported so far pertain to the solutions where the economy stays in the same regime (normal times or ELB) for all time periods. This begs for the following question: what happens for a switching solution where the economy is at the ELB in the short run and goes back to normal times before the shock is over? In that case, part of the increase in government expenses happens when the economy has left the ELB. As a result, one can expect the impact multiplier effect to depend crucially on $\ell$ as it can be shown that if that $\ell$ is large enough this economy spends a comparatively longer time at the ELB. In the following proposition, we describe how different the impact multiplier effect under a switching solution is compared to a non-switching solution. 


\begin{prop}
\label{prop:consumption_multiplier_mixed_solution}
Assume that assumption \ref{asp:pA_decay} holds and that
$\overline{d}(p)<d<\dbar(0)$. Let $\underline{\ell}$ be the integer
threshold from Proposition~\ref{prop:Mixed_solutions} below which the ELB
does not bind. Then the impact multiplier on consumption is such that
\begin{itemize}
    \item $m^{c,S}_{g}(\ell)=m^{c,N}_{g}(\ell)$ for
    $\ell=1,\dots,\underline{\ell}-1$.
    \item There exists a threshold $\ell^{+}$ such that
    $m^{c,S}_{g}(\ell)>0$ for all $\ell\geq\ell^{+}$.
    \item $\lim_{\ell\to\infty} m^{c,S}_{g}(\ell) = m^{c,L}_{g}(\infty)$ if and only if $p<\pbar$, otherwise it diverges to $+\infty$.
\end{itemize}
\end{prop}

\begin{proof}
See online Appendix, section F.    
\end{proof}

\begin{figure}[!htb]
  \centering
  \includegraphics[width=0.55\textwidth]{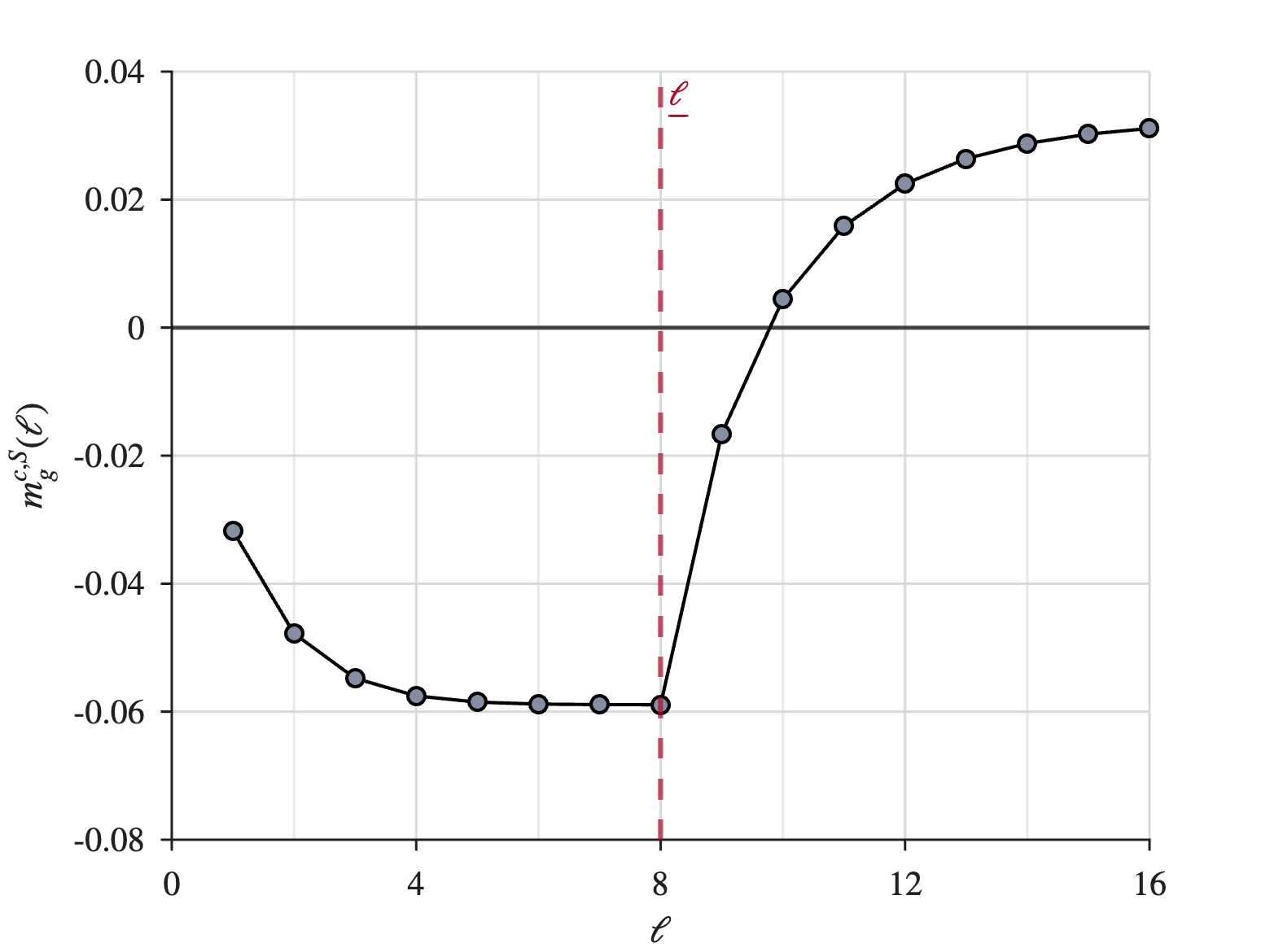}
  \caption{Illustration of the multiplier dynamics in the switching solution}
  \label{fig:switching_g}
\end{figure}

In words, for low values of $\ell$ the multiplier sequence behaves just as the one where the ELB never binds. When $\ell$ is sufficiently large however, the impact multiplier becomes positive. Eventually, as $\ell$ becomes arbitrarily large, the multiplier converges to the same sequence as in an ELB solution. We depict a typical sequence for a switching solution in Figure \ref{fig:switching_g}.

\begin{figure}[!htb]
  \centering  \includegraphics[width=0.55\textwidth]{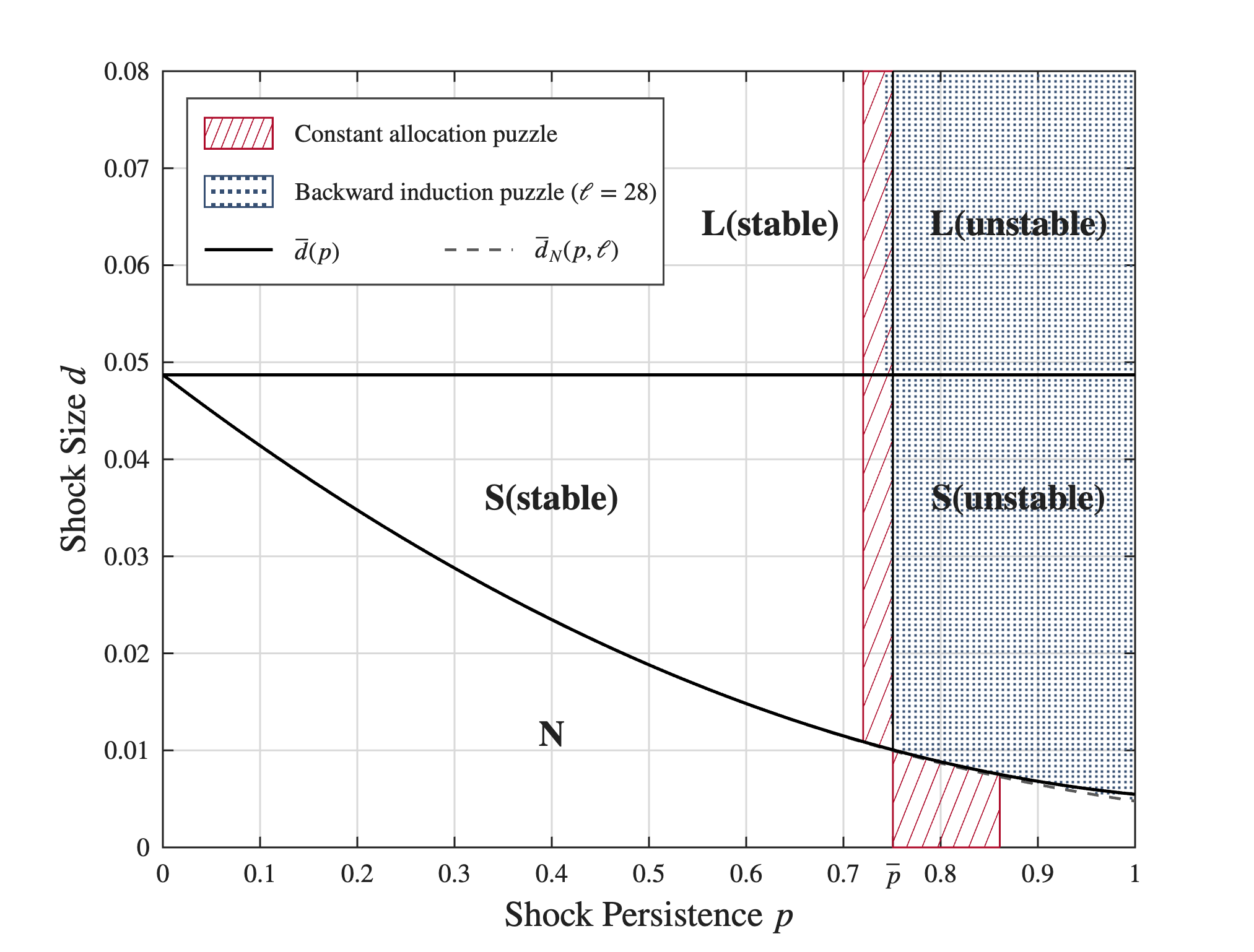}
  \caption{Puzzles with and without truncation}
  \label{fig:policy_puzzles}
\end{figure}

By and large, our conclusion at this point is that the graphical depictions that were introduced both  in \cite{eggertsson2011fiscal} and \cite{mertens2014fiscal} are extremely useful. The one in \cite{eggertsson2011fiscal} tells us what to expect in terms of multiplier effects when $\ell$ takes on a large value. The one in \cite{mertens2014fiscal} does not give us a multiplier that we can use, but instead it can tell us when the multiplier grows arbitrarily large with $\ell$. To expand on that insight, we finish by quantifying the parameter region for which our model displays a "policy puzzle". In a way that is similar in spirit to our Definition \ref{def:puzzle}, we state that the model displays a policy puzzle whenever the government spending multiplier at the ELB that is outside a reasonable empirical range. To construct this range, we borrow from \cite{Ramey2018government}, who report a $95\%$ confidence interval that spans the interval $[0,3.5]$ for the government spending multiplier on output. Using the same calibration as before, we compute the impact government spending multiplier on output using a standard absorbing Markov chain assumption alongside our truncated Markov chain where we assume a maximum duration of seven years, which is roughly the duration of the ELB episode that followed the U.S Great Recession. We also assume that the economy eventually reverts back to the intended steady state. We report these results in Figure \ref{fig:policy_puzzles}.

The main take-away from Figure \ref{fig:policy_puzzles} is that the policy puzzle region that we uncover with our shock specification is much larger than the one from the existing literature. For our calibration of $\ell$, the model displays essentially no policy puzzles to the left of $\overline{p}$. If we were to calibrate $\ell$ to realized ELB durations in Japan however, then our policy puzzle region would encompass the puzzle region from the existing literature to the left of $\overline{p}$. This is because our equilibrium construction converges to the one in the existing literature as $\ell\to\infty$ to the left of $\overline{p}$. Regarding the parameter region where $p>\overline{p}$, two features emerge. First, our construction uncovers no policy puzzles below $\overline{d}_N(p)$: this is because in that case the economy is in normal times and the multiplier falls in the empirically reasonable range. Second, the existing literature finds no policy puzzles above $\overline{d}(p)$: this is because in that case there is no MSV equilibrium using the standard procedure in the existing literature. There is thus no policy puzzle to discuss here because the model is silent. All in all, the main message from Figure \ref{fig:policy_puzzles} is that the potential for the New Keynesian model to produce policy puzzles may have been understated in the existing literature.

\section{Conclusion}
\label{sec:conclusion}

Using a truncated Markov chain, we have shown that one can study the unique path back to an arbitrary steady state in a standard New Keynesian model with an occasionally binding ELB. That shock structure helps uncover a number of insights about the results reported in \cite{eggertsson2011fiscal} and \cite{mertens2014fiscal}. The first one arises as a limit of our approach when persistence is low, but the second one doesn't. In that context, we have found that fiscal policy through higher expenditures unambiguously increases private consumption at the ELB. 

The simple framework that we have considered allowed us to prove formally a number of existence and uniqueness results. It also lead us to the conclusion that the range of parameters that injects a puzzle in the model is much larger than commonly believed. The natural question that arises then is: what would the prevalence of puzzles look like in a more general model that features endogenous persistence in the form of capital accumulation, consumption habits or inflation indexation? We are tackling this issue in ongoing work.

\bibliographystyle{apalike2}
\bibliography{CDLT_Morass_R1}

\setcounter{section}{0}
\renewcommand\thesection{\Alph{section}}
\renewcommand{\theequation}{\thesection.\arabic{equation}}
\setcounter{equation}{0}

\section{Appendix}
\label{sec:appendix}

\subsection{Proof of Proposition \ref{prop:pure_solutions}}
\label{proof:pure_solutions}
In Proposition 4 we characterize the two non-switching equilibrium paths: one that remains in the normal-time regime throughout the low-state episode and one that remains at the ELB. The main difficulty is that each candidate must satisfy its regime-verification condition in every low-state period for every finite maximum duration. For the normal-time candidate, the constant-allocation condition $\tilde\pi^N>\pibar$ is not always sufficient to establish validity at every finite horizon, because complex roots generate damped oscillations, causing the inflation path to fall below its limiting value for some finite $\ell$. We therefore study the entire inflation path and its lower envelope, which determines $\bar{d}_N(p)$. For the ELB candidate, the path is monotone as we solve backward, so validity reduces to checking the last possible low-state period. The eigenvalues of $(p\mathbf A^*)$ then determine whether the on-impact allocation converges or diverges as $\ell$ increases. The supporting results and detailed derivations are provided in Online Appendix H.4. 

\subsubsection{Normal-Time Regime}

Recall that the hypothetical normal-time equilibrium satisfies $\tilde Y_t=\mathbf A\mathbb E_tY_{t+1}+C_d d_t$. Under Assumption 1, backward induction gives the on-impact allocation
\begin{equation*}
    \tilde Y_t^N(\ell)=\left[\sum_{i=0}^{\ell-1}(p\mathbf A)^i\right]C_d d.
\end{equation*}
Let $V_2=[0,1]^\top$ and define
\begin{equation*}
    b_\ell(p)\equiv V_2^\top\left[\sum_{i=0}^{\ell-1}(p\mathbf A)^i\right]C_d,
\end{equation*}
so that $\tilde\pi_t^N(\ell)=d b_\ell(p)$. The Taylor principle guarantees that  $\rho(p\mathbf A)<1$, see Online Appendix H.2.3 for a formal proof. Hence,
\begin{equation*}
    \lim_{\ell\to\infty}\tilde Y_t^N(\ell)=(I-p\mathbf A)^{-1}C_d d\equiv\tilde Y^N.
\end{equation*}
In Online Appendix H.4.1 we show that
\begin{equation*}
b_\infty(p)\equiv\lim_{\ell\to\infty}b_\ell(p)=-\frac{\sigma\kappa(1+\eta\cbar)}{F_N(p)}<0,
\end{equation*}
where $F_N(p)\equiv(1-pm_{xx})(1-p\beta m_{\pi\pi})+\sigma\kappa(1+\eta\cbar)(\psi-pm_{x\pi})>0$. The limiting normal-time inflation rate therefore satisfies $\tilde\pi^N>\pibar$ if and only if
\begin{equation*}
d<\frac{\mu F_N(p)}{\sigma\psi\kappa(1+\eta\cbar)}=\bar{d}(p).
\end{equation*}

The limiting allocation alone does not always determine the validity of the finite-horizon normal-time path. In the Online Appendix H.2.2, we show that there exists a threshold $\psi^*>0$ such that the eigenvalues of $(p\mathbf A)$ are real when $\psi\leq\psi^*$ and form a complex-conjugate pair when $\psi>\psi^*$. Define $\underline b(p)\equiv\inf_{\ell\geq1}b_\ell(p)$. In the Online Appendix H.4.1 we establish that this is a negative number. With real roots, $b_\ell(p)$ decreases to $b_\infty(p)$, so $\underline b(p)=b_\infty(p)$. With complex roots, damped oscillations generate finite-horizon undershooting, and the infimum is attained with $\underline b(p)<b_\infty(p)$. Define
\begin{equation*}
    \bar{d}_N(p)\equiv\frac{\pibar}{\underline b(p)}.
\end{equation*}
If $d\in[0,\bar{d}_N(p))$, then, for every finite $\ell\geq1$,
\begin{equation*}
    \tilde{\pi}_t^N(\ell) = d b_\ell(p)\geq d\underline b(p)>\pibar.
\end{equation*}
Hence, the normal-time candidate is valid on impact for every finite maximum duration. This result also has a within-path interpretation. For any fixed $\ell$,
\begin{equation*}
    \tilde Y_{t+\ell-k}^N(\ell)=\tilde Y_t^N(k),
    \qquad
    k=1,\ldots,\ell.
\end{equation*}
It follows that the normal-time regime is selected throughout the $\ell$-period path. 

Moreover, the vector convergence above implies that on-impact private consumption and inflation converge to the finite values $\tilde c^N$ and $\tilde\pi^N$, respectively. 

The finite-horizon threshold $\bar d_N(p)$ is related to the constant-allocation threshold $\bar d(p)$ by
\begin{equation*}
    \bar{d}_N(p)=\bar{d}(p)\quad\text{when }\psi\leq\psi^*,
    \qquad
    \bar{d}_N(p)<\bar{d}(p)\quad\text{when }\psi>\psi^*.
\end{equation*}
Thus, the constant allocation characterizes the normal-time region with real roots, while finite-horizon undershooting narrows this region with complex roots. This proves the first statement.

\subsubsection{Validity and Asymptotic Behavior of the ELB Solution}

Recall that the hypothetical ELB equilibrium satisfies $\tilde Y_t^L=\mathbf A^*\mathbb E_tY_{t+1}+C_d^*d+E^*$. For a fixed maximum duration $\ell$, backward induction gives
\begin{equation*}
    \tilde Y_{t+\ell-k}^L(\ell)=\left[\sum_{i=0}^{k-1}(p\mathbf A^*)^i\right](C_d^*d+E^*),
    \qquad k=1,\ldots,\ell.
\end{equation*}
The matrix $(p\mathbf A^*)$ is element-wise nonnegative. When $d\geq\bar{d}(0)>\mu$, the forcing term $C_d^*d+E^*=\sigma(\mu-d)[1,\kappa(1+\eta\cbar)]^\top$ is element-wise negative. Hence,
\begin{equation*}
    \tilde Y_{t+\ell-(k+1)}^L(\ell)-\tilde Y_{t+\ell-k}^L(\ell)=(p\mathbf A^*)^k(C_d^*d+E^*)\leq0
\end{equation*}
for every $k=1,\ldots,\ell-1$. Thus, the allocation decreases weakly as we solve backward, and the highest inflation rate occurs in the last possible low-state period. Its verification condition is
\begin{equation*}
    \tilde\pi_{t+\ell-1}^L(\ell)=\sigma\kappa(1+\eta\cbar)(\mu-d)\leq\pibar
    \quad\Longleftrightarrow\quad
    d\geq\bar{d}(0).
\end{equation*}
Hence, if $d\geq\bar{d}(0)$, the ELB verification condition holds in every low-state period, and Proposition 3 establishes uniqueness. This proves the second statement. To characterize its limit, define
\begin{equation*}
    F_L(p)\equiv\det(I-p\mathbf A^*)=(1-pm_{xx})(1-p\beta m_{\pi\pi})-p\sigma\kappa(1+\eta\cbar)m_{x\pi}.
\end{equation*}
Recall from Proposition 1 that $\pbar\in(0,1)$ is the unique root of $F_L(p)=0$ for the parameter configurations considered in the paper. In the Online Appendix H.4.2, we derive the following limits:
\begin{equation*}
    \lim_{\ell\to\infty}\tilde Y_t^L(\ell)=
    \begin{cases}
        (I-p\mathbf A^*)^{-1}(C_d^*d+E^*)\equiv\tilde Y^L, & p<\pbar, \\
        -\infty, & p\geq\pbar,
    \end{cases}
\end{equation*}
where divergence is component-wise, linear at $p=\pbar$, and geometric for $p>\pbar$.
This proves the third statement and completes the proof.

\subsection{Proof of Proposition \ref{prop:Mixed_solutions}}
\label{proof:mixed_solutions}
Proposition 5 characterizes the intermediate shock region in which both non-switching constructions are not valid for every maximum duration. When $d<\bar{d}(0)$, the ELB cannot bind near the shock's end, whereas $d>\bar{d}(p)$ implies that hypothetical normal-time inflation eventually falls below $\pibar$. Hence, for sufficiently large $\ell$, the path contains both regimes. We show that, for $d\in(\bar{d}(p),\bar{d}(0))$, it switches only once: after leaving the ELB, the economy does not re-enter it before the shock ends. When $\bar{d}_N(p)<\bar{d}(p)$, paths in the remaining interval $d\in[\bar{d}_N(p),\bar{d}(p)]$ may switch more than once because of finite-horizon oscillations; we characterize them numerically. Proposition 1 gives the monotonicity of $\bar{d}(p)$ and the stability threshold $\pbar\in(0,1)$ for $(p\mathbf A^*)$; Online Appendix H.5 provides the remaining derivations.

\subsubsection{First Entry and the No-re-entry Property}
Fix a finite maximum duration $\ell$ and let $k=1,\ldots,\ell$ denote the backward step that determines period $t+\ell-k$. The hypothetical normal-time inflation rate is
\begin{equation*}
    \tilde\pi_{t+\ell-k}^N(\ell)=V_2^\top\left[\sum_{i=0}^{k-1}(p\mathbf A)^i\right]C_d d.
\end{equation*}
The right-hand side is independent of $\ell$. Schur stability implies convergence to $\tilde\pi^N$, while $d>\bar{d}(p)$ gives $\tilde\pi^N<\pibar$. At $k=1$, the normal-time candidate is valid because $d<\bar{d}(0)$. Define
\begin{equation*}
    \underline{\ell}\defeq\min\left\{k\geq1:V_2^\top\left[\sum_{i=0}^{k-1}(p\mathbf A)^i\right]C_d d\leq\pibar\right\}.
\end{equation*}
Convergence makes this set nonempty, whereas validity at $k=1$ implies $\underline{\ell}\geq2$. For every $k<\underline{\ell}$, the hypothetical normal-time candidate satisfies its verification condition and is selected. Hence, the selected path coincides with the hypothetical normal-time path through step $\underline{\ell}-1$. If $\ell<\underline{\ell}$, the economy remains in the normal-time regime. If $\ell\geq\underline{\ell}$, the normal-time candidate fails for the first time and the ELB is selected at $k=\underline{\ell}$.

Let $\tilde\pi_{t+\ell-k}^{SN}(\ell)$ denote the one-period normal-time candidate computed from the selected continuation allocation, and define $\chi_k\defeq\tilde\pi_{t+\ell-k}^{SN}(\ell)-\pibar$. Comparing it with the corresponding ELB candidate gives
\begin{equation*}
    \tilde\pi_{t+\ell-k}^{S}(\ell)-\pibar=
    \begin{cases}
        \chi_k, & \text{if }\chi_k>0, \\
        \left[1+\sigma\psi\kappa(1+\eta\cbar)\right]\chi_k, & \text{if }\chi_k\leq0.
    \end{cases}
\end{equation*}
Thus, $\chi_k>0$ selects normal time, while $\chi_k\leq0$ selects the ELB. No re-entry means $\chi_k\leq0$ for $k=\underline{\ell},\ldots,\ell$. By the definition of $\underline{\ell}$, $\chi_{\underline{\ell}-1}>0$ and $\chi_{\underline{\ell}}\leq0$.

For subsequent backward steps, the sign of $\chi_k$ cannot be determined directly from its second-order recursion because the two lagged terms enter with opposite signs. We therefore introduce an auxiliary sequence that reduces the second-order recursion to first order. Let $\lambda_1^*>\lambda_2^*>0$ denote the eigenvalues of $(p\mathbf A^*)$ and define $z_k\defeq\chi_k-\lambda_2^*\chi_{k-1}$. For $\ell\geq\underline{\ell}+1$, the recursion derived in Online Appendix H.5.3 gives the initial value
\begin{equation*}
    z_{\underline{\ell}+1}=\frac{\sigma\kappa(1+\eta\cbar)\left[\bar{d}(p)-d\right]}{1+\sigma\psi\kappa(1+\eta\cbar)}+\lambda_1^*\chi_{\underline{\ell}}-\frac{\lambda_1^*\lambda_2^*}{1+\sigma\psi\kappa(1+\eta\cbar)}\chi_{\underline{\ell}-1}<0.
\end{equation*}
The three terms are strictly negative, weakly negative, and strictly negative, respectively, because $d>\bar{d}(p)$, $\chi_{\underline{\ell}}\leq0$, and $\chi_{\underline{\ell}-1}>0$. Hence, $z_{\underline{\ell}+1}<0$. Since $\lambda_2^*>0$ and $\chi_{\underline{\ell}}\leq0$, we also have $\chi_{\underline{\ell}+1}=z_{\underline{\ell}+1}+\lambda_2^*\chi_{\underline{\ell}}<0$. Once the preceding two steps are in the ELB regime, $z_k$ satisfies the autonomous recursion
\begin{equation*}
    z_k=\frac{\sigma\kappa(1+\eta\cbar)\left[\bar{d}(p)-d\right]}{1+\sigma\psi\kappa(1+\eta\cbar)}+\lambda_1^*z_{k-1},
    \qquad k=\underline{\ell}+2,\ldots,\ell.
\end{equation*}
The negative constant and $\lambda_1^*>0$ imply that $z_{k-1}<0$ leads to $z_k<0$. Since $\chi_k=z_k+\lambda_2^*\chi_{k-1}$, the initial pair $z_{\underline{\ell}+1}<0$ and $\chi_{\underline{\ell}+1}<0$ propagates by induction. Hence, normal time is selected for $k<\underline{\ell}$ and the ELB for $k\geq\underline{\ell}$. In calendar time, the path switches once from the ELB regime to the normal-time regime, and the ELB binds on impact if and only if $\ell\geq\underline{\ell}$. This proves the first statement.

\subsubsection{Asymptotic Behavior}
For fixed $(p,d)$ and $\ell\geq\underline{\ell}$, the selected path is
\begin{equation*}
    \tilde Y_{t+\ell-k}^S(\ell)=
    \begin{cases}
        \left[\sum_{i=0}^{k-1}(p\mathbf A)^i\right]C_d d, & 1\leq k<\underline{\ell}, \\
        (p\mathbf A^*)^{k-\underline{\ell}+1}\tilde Y_{t+\ell-(\underline{\ell}-1)}^{S}(\ell)+\left[\sum_{i=0}^{k-\underline{\ell}}(p\mathbf A^*)^i\right](C_d^*d+E^*), & \underline{\ell}\leq k\leq\ell.
    \end{cases}
\end{equation*}
Here, we define $\tilde Y_{\mathrm{init}}$ as the continuation allocation inherited by the ELB block from the last normal-time backward step:
\begin{equation*}
    \tilde Y_{\mathrm{init}}\defeq\tilde Y_{t+\ell-(\underline{\ell}-1)}^{S}(\ell)=\left[\sum_{i=0}^{\underline{\ell}-2}(p\mathbf A)^i\right]C_d d.
\end{equation*}
By index symmetry, we have $\tilde Y_{t+\ell-k}^S(\ell)=\tilde Y_t^S(k)$; relabeling $k$ as $\ell$ yields
\begin{equation*}
    \tilde Y_t^S(\ell)=
    \begin{cases}
        \left[\sum_{i=0}^{\ell-1}(p\mathbf A)^i\right]C_d d, & \ell<\underline{\ell}, \\
        (p\mathbf A^*)^{\ell-\underline{\ell}+1}\tilde Y_{\mathrm{init}}+\left[\sum_{i=0}^{\ell-\underline{\ell}}(p\mathbf A^*)^i\right](C_d^*d+E^*), & \ell\geq\underline{\ell}.
    \end{cases}
\end{equation*}
Since $\underline{\ell}$ is fixed, only the second branch matters as $\ell\to\infty$. If $p<\pbar$, stability makes the inherited component vanish and the accumulated ELB component converge. Therefore,
\begin{equation*}
    \lim_{\ell\to\infty}\tilde Y_t^S(\ell)=(I-p\mathbf A^*)^{-1}(C_d^*d+E^*)\equiv\tilde Y^L.
\end{equation*}
This proves the second statement. If $p\geq\pbar$, iterating the recursion for $z_k$ to the on-impact value gives, for $\ell\geq\underline{\ell}+2$,
\begin{equation*}
    z_\ell=(\lambda_1^*)^{\ell-\underline{\ell}-1}z_{\underline{\ell}+1}+\frac{\sigma\kappa(1+\eta\cbar)\left[\bar{d}(p)-d\right]}{1+\sigma\psi\kappa(1+\eta\cbar)}\sum_{i=0}^{\ell-\underline{\ell}-2}(\lambda_1^*)^i.
\end{equation*}
Online Appendix H.2.4 gives $\lambda_1^*=1$ at $p=\pbar$ and $\lambda_1^*>1$ for $p>\pbar$. At $p=\pbar$, the negative constant accumulates linearly; for $p>\pbar$, the first term diverges geometrically. Thus, $z_\ell\to-\infty$. For $\ell\geq\underline{\ell}+2$, the no-re-entry result gives $\chi_{\ell-1}<0$, so $\chi_\ell=z_\ell+\lambda_2^*\chi_{\ell-1}<z_\ell\to-\infty$. The one-period identity yields $\lim_{\ell\to\infty}\tilde\pi_t^S(\ell)=-\infty$. This proves the third statement.

\end{document}